\documentstyle[eqsecnum,aps,epsfig]{revtex}
\def\Journal#1#2#3#4{{#1} {\bf #2}, #4 (#3)}
\def\NPA{{\em Nucl. Phys.} A}
\def\NPB{{\em Nucl. Phys.} B}
\def\PLB{{\em Phys. Lett.}  B}

\def\PRL{\em Phys. Rev. Lett}
\def\PRC{{\em Phys. Rev.} C}

\def\ZPC{{\em Z. Phys.} C}
\def\ZPA{{\em Z. Phys.} A}
\newcommand{\be}{\begin{equation}}
\newcommand{\ee}{\end{equation}}
\newcommand{\bea}{\begin{array}{c}}
\newcommand{\eaa}{\end{array}}
\newcommand{\ba}{\begin{eqnarray}}
\newcommand{\ea}{\end{eqnarray}}

\begin{document}
%
%
\date{\today}
\title{
Influence of Impact Parameter on Thermal Description
of Relativistic Heavy Ion Collisions at GSI/SIS.}
\author{Jean Cleymans$^1$, Helmut Oeschler$^2$
and Krzysztof Redlich$^{3,4}$}
\address{$^1$Department  of  Physics,  University of Cape Town,\\
Rondebosch 7701, South Africa\\
$^2$Institut f\"ur Kernphysik, Technische Universit\"at Darmstadt, \\
D-64289 Darmstadt, Germany\\
$^3$Gesellschaft f\"ur Schwerionenforschung, D-64291 Darmstadt, Germany\\
$^4$Institute for Theoretical Physics, University of Wroc\l aw,
PL-50204  Wroc\l aw, Poland\\}
\maketitle
\begin{abstract}
Attention is drawn to the role played by the size of the system
in   the  thermodynamic  analysis  of  particle yields in
 relativistic  heavy  ion
collisions at SIS energies. This manifests itself
in the non-linear dependence
of $K^+$ and $ K^-$  yields
in $AA$  collisions at 1 -- 2 $A\cdot$GeV
on the number  of participants.
It is shown that this dependence can be
quantitatively well  described   in  terms of a  thermal model
with a canonical strangeness conservation. The measured particle multiplicity
ratios ($\pi^+/p, \pi^-/\pi^+, d/p, K^+/\pi^+$ and $K^+/K^-$
but not $\eta/\pi^0$) in  central
Au-Au and Ni-Ni collisions at 0.8 -- 2.0 $A\cdot$GeV
are also explained         in the context of a thermal model with
a common freeze-out temperature and chemical potential.
Including the concept of collective flow a  consistent picture of
particle energy distributions is derived with the flow velocity
being strongly impact-parameter dependent.
\end{abstract}
\section{Introduction}
It  was  pointed out by Hagedorn \cite{hagedorn}
some thirty years ago that
thermal models overestimate the
production  of  anti-He$^3$ in proton-proton collisions
by  seven orders of magnitude when the grand canonical ensemble
is  used in its standard form
\cite{sonderegger}. The reason for this is that when
the  number  of particles as well as the interaction volume are
small  one has to take into account the fact
that the production of anti-He$^3$ must
be  accompanied  by the production of another three nucleons with energy $E_N$
in order  to  conserve the baryon number. Thus, the abundance will not
be
proportional to the standard Boltzmann factor given by
\begin{equation}
n_{\overline{He^3}} \sim
\exp \left( -{m_{\overline{He^3}}\over T} \right)
\end{equation}
but to
\begin{equation}
n_{\overline{He^3}} \sim
\exp \left( -{m_{\overline{He^3}}\over T} \right)
\left[ V \int {d^3p\over (2\pi)^3}
\exp \left( -{E_N\over T}\right)\right]^3
\end{equation}
since  three  additional  nucleons must be produced in order to
conserve the baryon number.
This   suppresses  the  rate  and  introduces  a  cubic  volume
dependence.
The  original  presentation of Hagedorn has been considerably
developed   and   expanded   in
refs.~\cite{redlich-hagedorn,old,Raf80,turko,mueller,derreth,gorenstein,cley1,cley3}.

Recently  it  has become clear that a similar treatment should be
followed  for  strangeness  production  in  the  GSI/SIS energy
range \cite{ceks}. This is not only due to the  fact that the
size  of the system is small but mainly because the temperature
is very low and particle numbers are small.
The abundance of $K^+$-mesons is then given by
\begin{equation}
n_{K^+} \sim
\exp \left( -{m_{K^+}\over T} \right)
\left[ g_{\overline{K}}V \int {d^3p\over (2\pi)^3}
\exp \left( -{E_{\overline{K}}\over T}\right)
+
g_{\Lambda}V \int {d^3p\over (2\pi)^3}
\exp \left( -{E_\Lambda\over T}\right)
\right] .
\end{equation}
since  the  strangeness must be balanced either by an anti-kaon
or  by a hyperon. $g_i$ are the degeneracy factors and $E_i$ the particle 
energies. This leads to a linear dependence
of the $K^+$ density on the size
of  the  system.  Such a dependence has
indeed been observed by
the KaoS collaboration for $K^+$-mesons \cite{MIS94,barth}.

In  this  paper  we would like to explore this idea in detail.
This volume dependence can now
be  tested for the first time by considering the  data
on   impact   parameter   dependence  which  are  now  becoming
available.
%
%

In  section  II we review the thermal
model with special emphasis on the canonical corrections due to
the exact conservation of quantum numbers.  In  section  III  we  present  a
comparison  with  the
experimental  data  from  SIS.
One part is devoted to a systematic study of central collisions
of various systems, another part to a detailed investigation of
the impact-parameter dependence. 
The last section is devoted to a
discussion of our results.
\section{Concepts and Predictions}
%
%
%
%
%
%
%
%
%
The exact treatment of quantum numbers in statistical mechanics
has been well established for some time now\cite{turko}.
It is in general obtained
by projecting the partition function onto the desired values
of the conserved charges by using the group theoretical
methods (for a  review see e.g. \cite{mueller}.)
For our purpose we shall only consider the conservation laws
related to the abelian U(1) symmetry group. In this case
the formalism is particularly simple and leads to the
following form of the canonical partition function
for a fixed  value of the conserved charge $Q$:
\begin{equation}
Z_Q={1\over {2\pi }}\int_0^{2\pi}d\phi e^{-iQ\phi }
{\tilde Z}(T,V,\phi )
\label{e:zq}
\end{equation}
\noindent
where $\tilde Z$ is obtained from the  grand canonical (GC) partition function
replacing the
fugacity parameter $\lambda_Q$  by the factor $e^{i\phi }$,
\begin{equation}
{\tilde Z}(T,V,\phi )\equiv
       Z^{GC}(T,V,\lambda_Q \to e^{i\phi })
\end{equation}
The particular form of the generating
function $\tilde Z$ in the above equation
is model dependent. Having in mind the applications of the statistical
description to particle production  in heavy ion collisions   we calculate
$\tilde Z$ in the ideal gas approximation, however, including
all particles and resonances
listed in \cite{pdg}.
This
is not an essential restriction, because, describing the freeze-out
conditions, we are dealing with a dilute system where the interactions
should not influence particle production anymore. We neglect
any medium effects on particle properties.
In general, however, already in the low-density limit, the modifications
of resonance width  or particle dispersion relation, in this particular
for $\Delta$ and $\pi$ \cite{nor,rus},
    cannot be excluded.
For the sake of simplicity, we use classical    statistics, i.e.~we 
assume temperature and density regime so that all particles
can be treated  using Boltzmann statistics.

In  nucleus-nucleus collisions  the absolute values
of baryon number, electric charge and strangeness are fixed by the initial
conditions.
Modeling particle production in statistical thermodynamics
would, in general, require the canonical formulation of all these
quantum numbers. 
>From the previous analysis \cite{ceks}, however, it is clear that
only strangeness should be treated exactly, whereas the
conservation of baryonic and electric charges can be described
by  the  appropriate chemical potentials in the grand canonical
ensemble.

Within the approximations described above and neglecting the contributions
from multi-strange baryons,
the generating function
in  equation  (\ref{e:zq})  has  the  following  form for a gas
having zero total strangeness, $S=0$:
\begin{equation}
Z_S(T,V,\mu_Q,\mu_B,\phi )=\exp (N_{s=0}+N_{s=1}e^{i\phi }+N_{s=-1}
e^{-i\phi })
\label{e:zs0}
\end{equation}
\noindent
where $N_{s=0,\pm 1}$ is defined as the sum over all particles and resonances
 having
strangeness $0,\pm 1$,
\begin{equation}
N_{s=0,\pm 1}=\sum_k Z_k^1
\label{e:N}
\end{equation}
\noindent
and  $Z^1_k$ is the one-particle partition function defined as
\begin{equation}
Z_k^1\equiv {{Vg_k}\over {2\pi^2}} \,
m_k^2 \, T \, K_2(m_k/T)  \, \exp (b_k\mu_B+q_k\mu_Q)
\label{e:zk1}
\end{equation}
\noindent
with the mass
$m_k$, spin-isospin degeneracy factor $g_k$,  the particle
baryon number $b_k$ and electric charge $q_k$. The
volume of the system is $V$ and the chemical potentials related with
the charge and the baryon number  are determined by
$\mu_Q$ and $\mu_B$ respectively.

With the particular form  of the generating function
equations (\ref{e:zs0},\ref{e:N},\ref{e:zk1})
the $\phi$-integration in equation (\ref{e:zq})  can be done analytically
giving the canonical partition function for a gas
with total strangeness $S$
in the following compact form \cite{cley1}:
\begin{equation}
Z_{S}(T,V,\mu_B,\mu_Q)=
Z_{0}(T,V,\mu_B,\mu_Q) I_S(x)
\end{equation}
\noindent
where $Z_0=\exp {(N_{S=0})}$ is the partition function of all particles
having zero strangeness and the argument of the Bessel function
\begin{equation}
x\equiv 2\sqrt {S_1S_{-1}}.
\label{e:x}
\end{equation}
\noindent
with $S_{\pm 1}\equiv N_{s=\pm 1}$.

The calculation of the particle density $n_k$
in the canonical formulation is straightforward. It amounts
to the replacement
\begin{equation}
Z_k^1 \mapsto \lambda_k \, Z_k^1
\end{equation}
of the corresponding one-particle partition function  in equation
(\ref{e:N})
and taking the derivative of the canonical (C) partition function
equation (\ref{e:zq})
with respect to the particle fugacity $\lambda_k$

\begin{equation}
n_k^C\equiv \lambda_k \left.{{\partial}\over
 {\partial\lambda_k}}\ln Z_Q(\lambda_k)\right|_{\lambda_k=1}
\end{equation}

As an example, we quote the result for the density of thermal kaons
in the canonical formulation assuming  the total strangeness of the
system $S=0$,
\begin{equation}
n_{K}^C={{Z_{K}^1}\over V}
{{S_1}\over {\sqrt {S_1S_{-1}}}}
{{I_{1}(x)}\over {I_0(x)}}.
\label{e:nkc1}
\end{equation}
Comparing the above formula with the result for thermal kaons
density in the grand canonical  ensemble,
$n_K^{GC}=(Z_K^1/V)\exp {(\mu_S/T)}$, one can see that the canonical
result can be obtained from the grand canonical one
replacing the  strangeness fugacity $\lambda_S \equiv
\exp {(\mu_S/T)}$ in the following way:
\begin{equation}
n_K^{C}=n_K^{GC}\left(\lambda_S\mapsto
{{S_1}\over {\sqrt {S_1S_{-1}}}}
{{I_1(x)}\over {I_0(x)}}\right).
\label{e:nkc2}
\end{equation}
In the thermodynamic limit both the canonical and the grand canonical
formulation are equivalent.
For a small  system, however, the
differences are large. This can be seen in the most transparent way
when comparing two limiting situations: the large and small
volume limit of equation (2.10)
In the thermodynamic limit $V\to \infty$ the argument
of the Bessel  function $x\to \infty $, thus
\begin{equation}
\lim_{x\to \infty } {{I_1(x)}\over {I_0(x)}}\to 1
\label{e:i1i01}
\end{equation}
\noindent
and the kaon     density is independent of the volume of the system
as expected in the grand canonical ensemble. On the other hand
in the limit of a small volume we have

\begin{equation}
\lim_{x\to 0 } {{I_1(x)}\over {I_0(x)}}\to {x\over 2}
\label{e:i1i02}
\end{equation}

\noindent
and  the particle density is linearly dependent
on the volume.
It is thus clear that the major difference
between the canonical     and the grand canonical
      treatment of the conservation laws appears
through different volume dependence of strange particle densities.
 The relevant parameter, $F_S$, which
 measures the deviations
 of particle multiplicities from their grand canonical result
is determined by the
 ratio of the Bessel functions
\begin{equation}
F_S\equiv  {{I_1(x)}\over {I_0(x)}}
\end{equation}
\noindent
with the argument $x$ defined in equation (\ref{e:x}).
In Fig.~1
we show the canonical suppression factor $F_S$ as a function
of the argument $x$.
 To relate the initial volume of the system
to the number of participants we use the approximate relation
$V\sim 1.9\pi A_{part}$.
The corresponding values of $x$ at SIS, AGS and SPS energies are
calculated with the     baryochemical potential
and temperature extracted from the measured particle multiplicity ratios.
The results in Fig.~1 show  the importance of the canonical
treatment of strangeness conservation at SIS energies. Here,
the canonical suppression factor can be even larger  than
 an order of magnitude.
For central Au-Au collisions at AGS or SPS energies this suppression
is not  relevant any more and the (GC)-formalism is adequate.
In general, one expects that the statistical interpretation of
particle production in heavy ion collisions requires the
 canonical treatment of strangeness
conservation if the CMS collation energy $\sqrt {s} <2-3$ GeV.
This is mainly because at these energies the freeze-out
temperature is still too low to maintain large-argument expansion
of the Bessel functions in equation (\ref{e:i1i01}).

At low temperatures one needs to take into account
the width of resonances.
This  is  because
the number of pions coming from e.g.~the  
decay of a $\Delta$-resonance is increased by the width of
the  $\Delta$.
The approximation
of the width by a delta function is
therefore not  justified  since
an appreciable number of particles
come from the decay of resonances below the resonance mass.
One therefore  replaces the
one-particle  partition  function  in  equations (\ref{e:N}) and
(\ref{e:zk1})
by:
\begin{equation}
Z_R^1=N \, {{Vd_R}\over {2\pi^2}} \,
T \, \exp (b_k\mu_B+q_k\mu_Q) \,
\int_{s_{min}}^{s_{max}} ds \, s \, K_2(\sqrt {s}/T) \,
{1\over \pi} \, 
{{m_R\Gamma_R}\over
{    (s-m_R^2)^2 + m_R^2\Gamma_R^2 }}
\end{equation}
\noindent
where $s_{min}$ is chosen to be the threshold value for resonance
decay and $\sqrt {s_{max}}\sim m_R +\Gamma_R$. The normalization constant
$N$ is adjusted so that the integral over the Breit-Wigner factor
gives one.

Within the above model,
the particle densities depend on four parameters:
 the  chemical potentials, $\mu_Q$ and $\mu_B$,
related with the   (GC)-description of charge and baryon number
conservation, the temperature $T$ and the initial volume of the system
appearing through the   canonical treatment of strangeness conservation.
Constraints on these variables arise from the
isospin
 asymmetry  measured  by the baryon number divided by twice the
 charge,  $B/2Q$. For an isospin symmetric system this ratio is
 simply 1, for Ni+Ni it is 1.04 while for Au+Au this ratio is 1.25.

We are thus left with three independent parameters.
For simplicity,
the volume $V$ will be identified with the         volume
of the system created initially in $AA$ collisions 
estimated from the atomic number of colliding nuclei
and from the impact parameter by
 using  geometric arguments. In particular we  use
 the relation of the volume parameter and the number of participating
nucleons as it was indicated in Fig.~1. 

In the following section we will discuss to what extent the
thermal  model can be used to understand
particle production in nucleus-nucleus collisions at SIS energies.
\section{Comparison with Experimental results}
This section is divided into three parts. In the first one
we discuss the general trends found in thermal models.
In particular
we illustrate the sensitivity of  particle ratios on the
temperature, $T$, and on the baryon chemical potential, $\mu_B$.
In
the second part we discuss
particle ratios measured in
central collisions and compare experimental results obtained at
SIS with the expectations of the thermal model.
This avoids the problem that different particle species might
originate  from different impact parameter regimes as it is the
case with
inclusive studies.
The impact parameter dependence is studied in the third part of
this section.
\subsection{General Trends}
>From equations (\ref{e:nkc1},\ref{e:i1i01},\ref{e:i1i02}) one sees
that  at fixed
temperature and chemical potentials the         volume dependence
of kaon multiplicity in the canonical  and
the grand canonical ensemble
is as following:
\begin{equation}
\left<N_K\right> \sim
\left\{ \begin{array}{r@{\quad :\quad}l}
V~  & V\mapsto \infty ~~~({\rm GC})\\
V^2 & V\mapsto 0 ~~~~~~({\rm C}).
\end{array} \right .
\end{equation}
This  effect  is shown explicitly in
Fig.~2
where  the different particle ratios are
calculated  as a function of the radius
of the system.  In order to make the analysis more complete
one needs, in addition,  to take into account the contribution of resonance
decays to particle production. We use all the known branching
ratios as given in \cite{pdg} to
calculate the particle multiplicities shown
in Fig.~2.

The ratio $K^+/\pi^+$ involves  one strange
particle, and  according to equation (\ref{e:nkc1})  should show 
substantial dependents on
volume. Indeed, it
increases smoothly from zero up to the
value  given  by  the grand canonical ensemble. The increase is
faster  for higher  temperature   because
the        value of
the argument $x$ of the Bessel functions in equation (\ref{e:nkc1})
increases.
In the $K^+/K^-$ ratio two strange particles are involved and
the  volume effect  cancels out exactly.
However,  for  very small values of the volume, there are a few
non-strange  resonances  that  decay  into kaon pairs or into a
kaon and a hyperon, this leads to the sharp rise in the value of
this  ratio for  increasing  volumes  before it flattens off and
becomes volume independent.
The  ratio  $\phi/\pi^+$ involves only non-strange
particles  and therefore
         is independent of the size
of  the system. The same behavior is expected for $\eta$ mesons which
is treated as non-strange particle, too.

 Figure~3  evidences the locations of the freeze-out temperature $T_f$ and 
of the freeze-out $\mu_B^f$
yielding the
various particle ratios. This figure shows also the sensitivity
when varying  their values           within the limits occurring in
the experimental results. Discussing these trends requires different
treatment of strange and non-strange particles.

The interesting feature of the deuteron to proton ratio, $d/p$,
and the pion to proton ratio,
$\pi^+/p$, is that they allow a good determination of the range
 of the thermal parameters. The $\pi^+/p$ curve
in the $(T-\mu_B)$ plane
 shows a  temperature saturation
for large $\mu_B$  which establishes
the upper limit of the freeze-out temperature $T_f$.
On the other hand,  the $d/p$ ratio
fixes   the  range of the freeze-out $\mu_B^f$ as
 it shows a steep dependence
on  the temperature. In addition, as seen in Fig.~3,
the variations of 20-30$\%$
on the value of
 the $d/p$ ratio
give a very similar range in the $(T-\mu_B)$ plane, making
this observable particularly useful for the extraction
of the   freeze-out parameters.

 In the  SIS-energy
range one expects, in the thermal model,
a different dependence of the strange and non-strange
particle yields on the volume of the system.
Consequently, the ratio of
 strange to non-strange particle  leads
to a strong variation of thermal parameters with the system size.
In Fig.~4 we show the $K^+/\pi^+$ ratio
for different volumes. As expected, changing the volume
implies a substantial modification of the curve in the $(T-\mu_B)$ plane
corresponding to a fixed value of the $K^+/\pi^+$ yields.
Thus, calculating the
strange to non-strange particle ratio requires additional care
of the system size. In our approach, however, the volume
  is not treated as an additional parameter but is rather
       related  with  the number of participating
nucleons in $AA$ collisions.
For   a   given  system  size  the  $K^+/\pi^+$  ratio  clearly
determines
the lower limit of the freeze-out temperature. 
The curves in Fig.~3 are calculated for $R$ = 4 fm.
The $K^+/K^-$ in Fig.~3
yield exhibits similar behavior as $d/p$,    i.e.   showing
a  very  strong dependence on the temperature but a rather weak
dependence on the baryon chemical potential, $\mu_B$.
This lines do not depend on the volume.

An analysis of particle production in heavy ion collisions
within a thermal fireball model requires  two experimentally measured
ratios to fix the freeze-out parameters, $T_f,\mu_B^f$, and
knowledge of the number
of participating nucleons, $A_{part}$ to establish the
size of the fireball.
The knowledge of more particle ratios allows to test the concept of
a unique freeze-out time.
In the following  section we compare
the predictions of the thermal
model with experimental results for central $AA$ collisions
at SIS energies.
\subsection{Central collisions}
In central collisions, the number of participating nucleons is
maximal.
However, experimental results for
zero impact parameter, $b$ = 0, are not directly measured.
However, for many experiments good-quality
impact-parameter  results are available and
an extrapolation to $b = 0$ can be performed.
The results of this extrapolation are summarized in Table 1.
We discuss below the different entries in this table.

The results in Table 1 for pions are
obtained from \cite{MUE97,PEL97,PEL97a}.
For the ratio $\pi^+/p$ we used  results
from inclusive measurements since it
was established that
the pion multiplicity divided by the number of participants,
$M_{\pi}/A_{part}$, does not vary with $A_{part}$
\cite{MUE97,PEL97,PEL97a,HAR87}.
At Ni-Ni at 0.8 $A\cdot$GeV $\pi^-$ data are not available and 
$\pi^+ = \pi^-/1.2$ has been used to account for the isospin asymmetry.
%
The $K^+$ results for
Ni+Ni are from \cite{barth} which are in very good
agreement with \cite{best}. The $K^+$ yield rises strongly with
centrality as shown in \cite{barth}.
At 1.0 and 1.8 $A\cdot$GeV the $K^+/\pi^+$ ratio was obtained by extrapolating
to $b=0$. 
At 0.8 $A\cdot$GeV we scaled with
the inclusive $K^+/\pi^+$ ratio between 1.0 and 0.8 $A\cdot$GeV.
The impact-parameter dependence for $K^-$ and $K^+$
is nearly identical \cite{barth} and we used therefore the $K^+/K^-$
ratio of the inclusive measurements.
The multiplicity of $K^+$ divided by the number of participants,
$M_{K^+}/A_{part}$, in Au+Au
increases strongly with impact-parameter
as shown in \cite{MIS94,MANG97}.
The $\pi^-$/$\pi^+$ ratios  show -- if at all -- 
only a very slight increase with
centrality \cite{PEL97a,WAG98} and Table 1
summarizes the results for inclusive studies.
These values in \cite{PEL97a,WAG98} agree
very well with the isobar model.
%
The ratios of $d/p$ obtained in Ni+Ni are from
\cite{HON98} and those for Au+Au have been taken from \cite{Speer}.
%
The ratios $\eta/\pi^0$ for Ni+Ni are from \cite{AVE98,MAR97}.
At 1 $A\cdot$GeV Ni+Ni has not been measured, yet Ar+Ca and Kr+Zr
yield the equal ratios taking then also for Ni+Ni, at 0.8 $A\cdot$GeV
only Ar+Ca has been studied and this value is given in Table 1.
The value at 1.8 $A\cdot$GeV has been obtained by interpolation between
1.0 and 1.93 $A\cdot$GeV.
For Au+Au the results from \cite{AVE98} have been corrected
for the increase with centrality according to \cite{WOL98}.
%

Figure~5 shows the lines in the $(T-\mu_B)$ plane corresponding
to the measured particle ratios in  Au-Au collisions at 1 $A\cdot$GeV.
The experimental errors are for simplicity
not shown in the figure.
All lines, except  the one for $\eta /\pi^0$,
have a common
crossing point  around $T\sim 50$ and
$\mu_B\sim 822$ MeV. A value of $R\sim 6.2$ fm
is needed to describe      the measured $K^+/\pi^+$ ratio with the
freeze-out parameters
 extracted for $\pi^+/p$, $\pi^+/\pi^-$ and $d/p$.
This radius corresponds to $A_{part}\sim 330$ participating nucleons and is
compatible with the one expected for central Au-Au collision.

The strangeness suppression due to the canonical treatment of the
conservation laws is very clear in the comparison of the thermal
model with the  Au-Au 1 $A\cdot$GeV data.
Using the {\it grand
canonical} formulation of the strangeness conservation
 one would get the value   $K^+/\pi^+\sim 0.04$
 which
overestimates the data by more than an order of
magnitude. This  shows that the conditions for thermal
particle phase space at SIS energies are far
from the grand canonical limit.

In Fig.~6 we
show the corresponding
results in the $(T-\mu_B)$ plane for Ni-Ni collisions at three
different
incident energies: 1.8, 1.0 and  0.8 $A\cdot$GeV.
As     for  the    Au-Au     data     we    see    in
Fig.~5 that all
particle ratios, besides $\eta/\pi^0$ can be described by the
same values of the freeze-out parameters.
We notice the smaller
radius of $\sim 4$ fm which is  compatible with the smaller size of Ni.
The measured $K^+/K^-$ ratio leads to a band
in the $(T-\mu_B)$ plane which barely misses
the common crossing point for  the mean values of
$\pi^+/p$, $d/p$ and $K^+/\pi^+$.
However, taking into account the
experimental errors on the above particle ratios leads to a common,
narrow $(T-\mu_B)$-band which  contains also the line corresponding
to the upper experimental limit for $K^+/K^-\sim 30$.
Allowing for a  drop in the  $K^-$ mass as proposed in 
model calculations
\cite{schaffner,brown,weise} leads to a  shift of the $K^+/K^-$ band
towards the
left in Fig.~6, thus  leading to a better agreement
with the other data.
New experimental results on the $K^+/K^-$ ratio also for other collision
systems are needed to clarify this open question.

 The values for the  of  $d/p$  ratio
shown in Fig.~6
for Ni-Ni collisions at 0.8 $A\cdot$GeV
were obtained by extrapolating the experimental
measurements at E/A=1.06, 1.45, 1.93 summarized in Table 1 using
a linear and a polynomial fit giving the value $d/p$ = 0.4, 0.43.
As one can see the 10$\%$ deviation
on $d/p$ ratio
does not modify substantially the freeze-out line
 in the $(T-\mu_B)$ plane.

The results of Figs.~5, 6
show   that for central Au-Au and Ni-Ni collisions the particle ratios,
    $\pi^+/p$, $K^+/\pi^+$, $\pi^+/\pi^-$, $K^+/K^-$ and $d/p$,
 lead to a common crossing point in the  $(T-\mu_B)$ plane.
  The appearance of the common freeze-out for all
these  particles  is  strong  support  for  chemical
equilibrium of these particles in the  thermal model.

In Fig.~7 we show the results for
the freeze-out temperature
and chemical potential corresponding to different incident energies.
It can be seen that
the freeze-out temperature  $T_f$ increases with E/A
whereas $\mu_B^f$ shows the opposite trend.
 In the energy range
considered  this  dependence  can be approximated by a straight
line.
The freeze-out parameters are also seen in Fig.~7 to be
 different in Ni-Ni and
Au-Au collisions even for the same incident energy.
It is interesting to remark that at a given incident energy the extracted values
of $T_f$ are smaller for central Au-Au collisions than for central Ni-Ni ones.  

Using the freeze-out parameters  relevant for SIS energies
we can compare the results with the previous findings
for AGS \cite{stachel,elliot} and
 SPS \cite{sollfrank,stock,jaipur,marek}
energies. The chemical freeze-out points in the
$(T-\mu_B)$ plane for relativistic
heavy ion collisions are  shown in Fig.~8
together with the  predictions
for $e^+e^-$ and p-p collisions \cite{bec}.
Connecting these points leads
to a unique freeze-out curve in the $(T-\mu_B)$ plane \cite{Hag80,peter}.
It has been shown~\cite{prl} that
 the energy per hadron along the freeze-out  curve
 (before the final decay of the hadron) is approximately 1 GeV.

It is a common feature of Figs.~5 and 6 that the freeze-out line for
$\eta$ production does not cross the common chemical freeze out extracted
for all the other particle species. This deviation does not arise from the
selection of central collisions. Only for Au-Au collisions we have
chosen the $\eta/\pi^0$ ratio of central collisions. In the Ni-Ni
system -- due to the lack of experimental results -- inclusive
values are given. For central Ni-Ni collisions the discrepancy would
therefore increase. The results in \cite{WOL98} show that the $\eta$ yield
rises more than linearly with $A_{part}$. This disagrees also 
with the expectations
from the thermal concept discussed along with Fig.~2.

The problem of      $\eta$  abundance in the thermal model
could have several origins. One possibility would be due to the
sequential freeze-out for different particle species.
Here,            $\eta$  could be produced earlier with higher temperature,
roughly corresponding to the value obtained from the
transverse momentum slope parameter of 80 MeV. In this case,
as seen in Fig.~6, the large value
of the $\eta /\pi^0$ ratio can
be understood. 
In recent works \cite{WAG98,WAG99,oeschler} experimental 
evidence is given that high-energy
pions are emitted earlier than those with lower energies.
This points towards a span in freeze-out times but does not affect
the particle ratios which are dominated by low-energy pions.
The concept of sequential freeze-out, however,
might lead to a problem with the interpretation of
all other particle
ratios which are otherwise well described by the model.
We stress that
the same thermal model when  applied to  AGS and SPS energies explains
particle productions there with a single set of freeze-out parameters
for
all particles including $\eta$ at SPS energies. It seems rather difficult to
argue the appearance of a sequential freeze-out at SIS and its
absence at AGS and SPS energies.

The discrepancy of the thermal model on the level of
$\eta$ production could  possibly be  related to the hidden
strangeness content of the $\eta$  meson.
The crucial  difference in the
thermal model interpretation of particle production at SPS, AGS and
SIS energies is due to the  canonical strangeness suppression.
In our formalism the     $\eta$  is
considered as a non-strange particle.
It is conceivable
that due to hidden strangeness there should be   corrections to
the $\eta$ yield  from the canonical strangeness  conservation. We do not
know, yet, however, how these corrections
 could be included in a consistent way.  
The same arguments would apply also for the production of $\phi$. Yet, 
$\phi/K^-$ (inclusive) seems to fit into the common crossing shown for
Ni-Ni at 1.8 $A\cdot$GeV.
The large $\eta$  yield could
 well be of
    dynamical origin which apriori cannot be explained by the thermal
freeze-out model applied in this work.
Thus, we leave
the discrepancy of $\eta$  production as an unsolved puzzle in the
thermal model.


Concluding this part, we find that the thermal model
gives a consistent description of both strange and
non-strange
particle production in central nucleus - nucleus collisions at SIS
energies with the exception of the $\eta /\pi^0$ ratio.
There is a serious problem with the understanding
of this ratio in terms of the thermal model.
Theory and experiment deviate by a factor of 2 to 8, depending on
the incident energy. The same problem
has been indicated recently in the context
of dynamical model for pion and $\eta$  production \cite{nor}.
We would like to point out that recent transport-model calculations
reproduce the measured $\eta$ and $\pi^0$ spectra \cite{brat98}. 
We would also like to draw attention to the systematics introduced by V. Metag
in which pions and $\eta$'s exhibit a common trend while $K^+$, $K^-$ and
$\phi$ show lower yields \cite{Metag}. 

To check further the consistency
of the model with the experimental data
the impact parameter dependence of particle production and particle spectra
are discussed in the next section.
%
%



%
\subsection{Impact Parameter Dependence}
As  already mentioned, the multiplicity of $K^+$ divided by the
number of participants, $K^+/A_{part}$, increases
strongly with the centrality while the corresponding ratio for
pions, $K^+/A_{part}$, is
constant \cite{barth}. Consequently,   the pion yield is proportional
to the number of nucleons in the initially created fireball while
the multiplicity of $K^+$  scale with $A_{part}$
like $K^+\sim A_{part}^\alpha$ with $\alpha >1$.
The experimental
results on $K^+/\pi^+$ and $K^+/A_{part}$ ratios in Au-Au collisions
  from Ref. \cite{MANG97}
are  shown  in  Fig.~9.
The canonical treatment of strangeness conservation predicts
the yield of strange particles to increase quadratically
with the  number of participants
(see eq. (2.14)).
An additional complication
is due to
the possible variation of  the  freeze-out parameters with centrality.

In Fig.~9 the dashed-line
describes the results of the
thermal model under the assumption that both $T_f$ and $\mu_B^f$
are $A_{part}$ independent.
One sees   that
already under this simple approximation the
 agreement of the model and the experimental data is very
satisfactory.
 The small deviations between  the model and the data
 can be accounted    for by
the variation of the freeze-out parameters with $A_{part}$.

We have calculated the possible change of $T_f$ and $\mu_B^f$
with $A_{part}$ from the experimental data on $K^+/\pi^+$
and $d/p$ ratios     measured for
 two different values of $A_{part}$
(see Fig.~10). The resulting
freeze-out parameters are shown in Fig.~11. One can see
that the variation of $T_f$ and $\mu_B^f$ with $A_{part}$ are
small.
It is interesting to note that peripheral collisions yield higher
temperatures than central collisions. The results from central $Ni-Ni$ 
collisions fit perfectly into the trend with $A_{part}$.

 In Fig.~9 the
dashed-dotted lines were obtained parameterizing the $T_f$ and
$\mu_B^f$  dependence  on  $A_{part}$
using the small variation in freeze-out parameters
obtained  from  Fig.~11.
The shape of the experimental data is now well reproduced.

The rather small variation of the freeze-out temperature
with impact parameter
shown  in  Fig.~11  comes as a  surprise
since the experimental results
on the apparent inverse slope parameter $T_{app}$ of particle yields
shows a strong
dependence on $A_{part}$. The freeze-out
temperature of $\sim 50$ MeV derived from our analysis is also substantially
lower than the value previously  obtained from  the particle spectra
\cite{MUE97,PEL97,PEL97a,HON98,oeschler,herrmann,TAPS}.

 We now turn our attention
to the differential cross sections  for particle production.
In  Fig.~12 we show the variation of the inverse
slope parameters observed for  various particle species \cite{MUE97,HON98}.
Yet, adopting our result, that there exists a common,
 impact-parameter independent,  freeze-out
temperature for all particles one
needs to show that the variation in slope results from radial
flow varying with impact parameter. We extract the values
of the corresponding flow parameters $\beta$ using the
Siemens-Rasmussen formula \cite{siemens}.
The   resulting   values  for  the  flow  velocity,
$\beta$, are
summarized in  Fig.~12.
 With the  freeze-out
temperature $T_f\sim 50$ MeV the spectra of $p$, $K^+$ and
$\pi^+$ are, within the experimental uncertainty,     well
 explained with the same $\beta$ as shown in Fig.~12.
The variation of $\beta$  with impact parameter turns out to be
very large.

The $A_{part}$ dependence of various particle yields
($Y \sim A_{part}^{\alpha}$) has been discussed in \cite{MUE97} and
the exponent $\alpha$ has been suggested to be related with the difference in
total energy needed for the production of the studied particle
and the one available in $NN$ collisions. In the presented concept
the exponent $\alpha$ is only due to volume-dependent strangeness suppression.
In this picture the exponent for $K^+$ and for $K^-$ production
is predicted to be equal while in the frame of the energy argument \cite{MUE97}
$\alpha$ should be higher for $K^-$ than for $K^+$ due to different
production thresholds (neglecting in-medium mass modifications).
It is clear that any non-isotropic emission pattern is beyond the scope of
this model. However, -- as naively expected -- the slopes of $K^+$ and of 
$K^-$ do not have to be equal as different resonance decays contribute.

The results of this section show  that
the thermal model gives  a consistent
picture  for  particle  production  in  $AA$  collisions  at SIS
energies.
In addition to the correct predictions for particle yields
the model can also explain  the $A_{part}$ dependence
of $K^+$ cross section. Assuming radial flow
as an origin of the shape of particle spectra one can understand
quantitatively the inverse slopes with a common temperature and
radial flow velocity for all particles.
The chemical and thermal freeze-out seem to be very close  at SIS energies.
This  is because the particle multiplicities and momentum spectra
can be
explained with the same temperature.
\section{Summary}
In summary, the thermal model provides a remarkably consistent
description  of  the  experimental  data in the GSI/SIS energy
range.  The  abundances of particles, $K^+, K^-, p ,d, \pi^+ $ and
$\pi^-$  (with  the  notable  exception of $\eta$'s) seem to
come  from a common hot source with a surprisingly well defined
temperature,  $T\approx 50,54,70$ MeV, and
baryon chemical potential $\mu_B\approx 825,805,750$ MeV for central
Ni-Ni at 0.8, 1.0, 1.8 $A\cdot$GeV  and
correspondingly  $T\approx 52$ MeV and
$\mu_B\approx 822$ MeV for central Au-Au collisions at 1.0 $A\cdot$GeV,
 as can
be seen clearly from Figs.~5,7 and  8.
These temperatures  are          lower  than  the  ones observed in the
particle  spectra  but  here  again  a  common  explanation is
possible  in  terms  of  hydrodynamic  flow. Flow differentiates
between  heavy  and  light  particles  since they  acquire
the same boost in
velocity   but  of course very different kicks in momenta.
This
is clearly seen in the GSI/SIS data shown in Fig.~12
which also summarize the transverse flow velocity for pions, kaons and
protons.

The common
 crossing points exhibited in Figs.~5, 6
 are a very strong argument for
 chemical equilibrium. The deviations found for the production of $\eta$ mesons
 clearly ask for an explanation.
%
%
\section{Acknowledgments}

We acknowledge stimulating discussions with R. Averbeck, P. Braun-Munzinger,
B. Friman, V. Metag, W. N\"orenberg and W. Weinhold.
\newpage
%
%
%
\begin{table}
\begin{tabular}{|l l| l l c c c l|}
\hline
Reaction &E & $\pi^+$/p & K$^+$/$\pi^+$ &
$\pi^-$/$\pi^+$ & d/p$$
& K$^+$/K$^-$$^a$ & $\eta/\pi^0$\\
 &  $A\cdot$GeV  &&&&&&\\
\hline
Ni+Ni & 0.8 & 0.05 & 0.0003  &  &   & & {\it 0.004}\\
Ni+Ni & 1.0 & 0.08  & 0.001 & {\it 1.2} & 0.37 &
& {\it 0.013}\\
Ni+Ni & 1.8 & 0.17& 0.0084 & {\it 1.05} & 0.28 & 21$\pm$9
&{\it 0.03}\\
\hline
Au+Au & 1.0 & 0.052 & 0.003 & 2.05{\it (1.94)} & & & 0.03/{\it 0.014}\\
\hline
\end{tabular}
\caption{{\it Experimental results for different particle ratios in
central $AA$ collisions.
Values in italics are for inclusive measurements. The errors are estimated to
20 -- 30 \%.}}
\label{exp_ratio}
\end{table}
%
%
%
%
%
%
%
%
%
%
\newpage
\newpage
\begin{figure}
\begin{center}
\vspace{1.5cm}
\epsfig{file=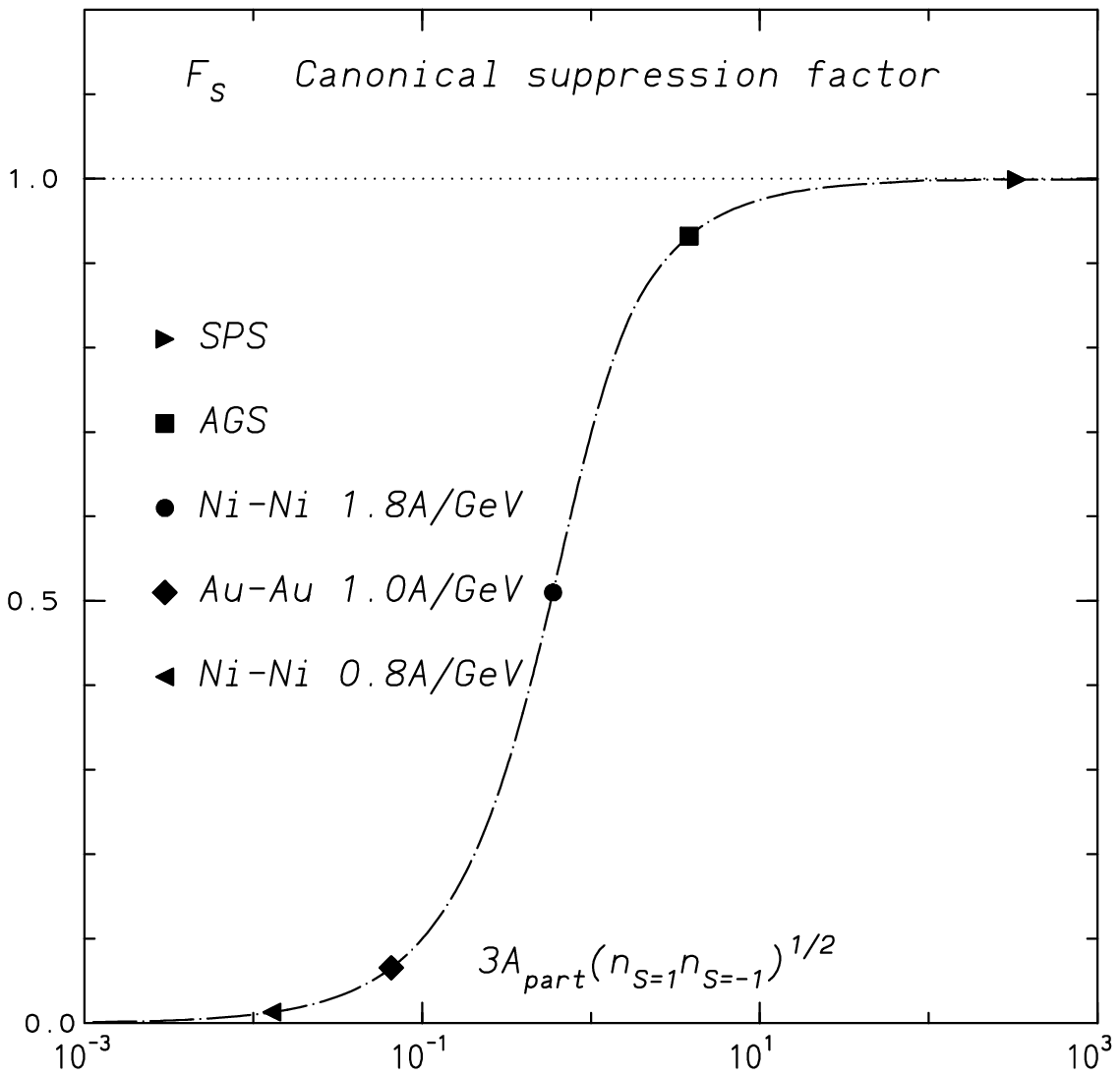,width=6.5cm}
\end{center}
\end{figure}
\newpage
%
%
\begin{figure} 
\begin{center}
\epsfig{file=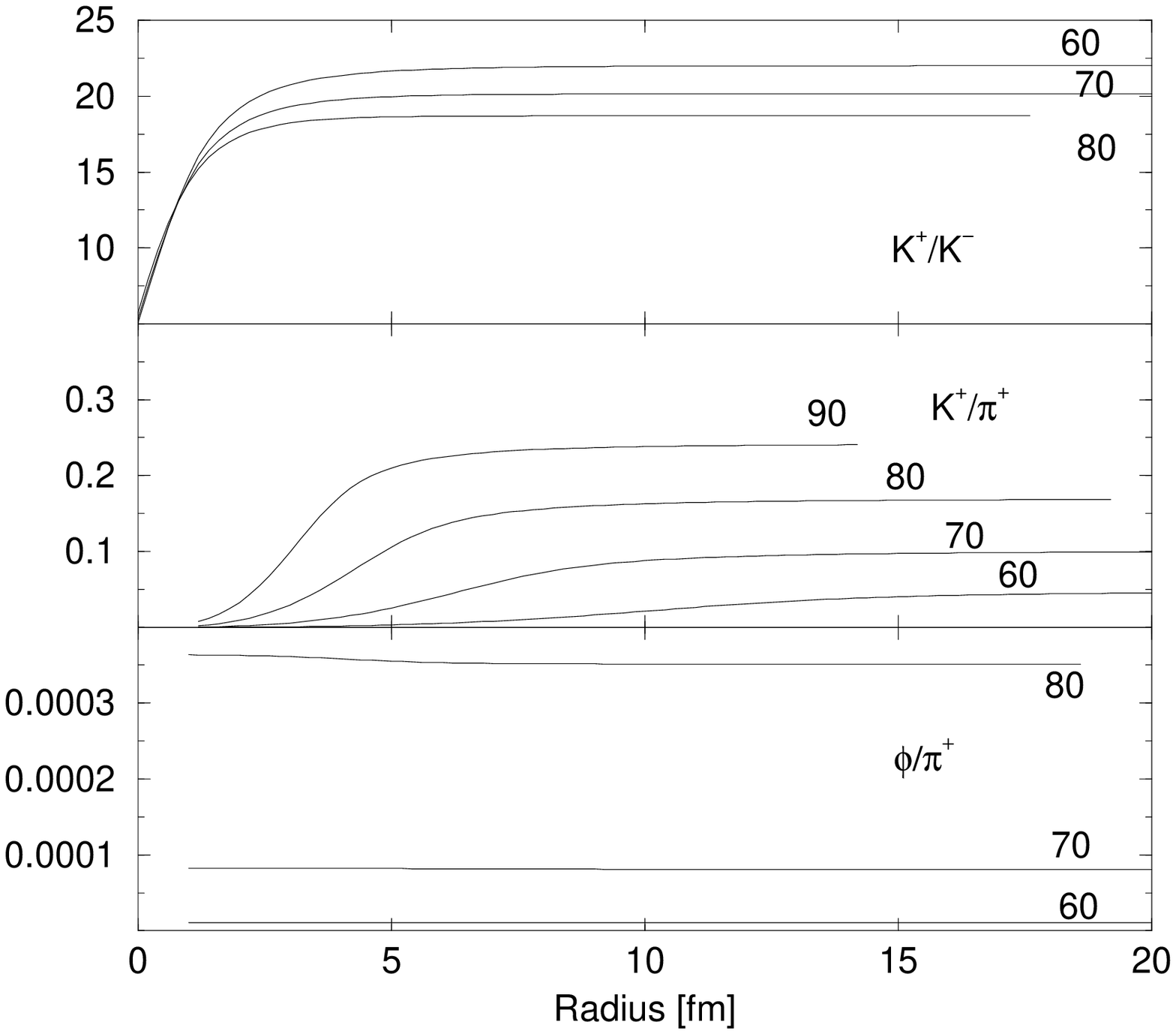,width=7cm}
\end{center}
\end{figure}
\newpage
%
%
\begin{figure} 
\begin{center}
\epsfig{file=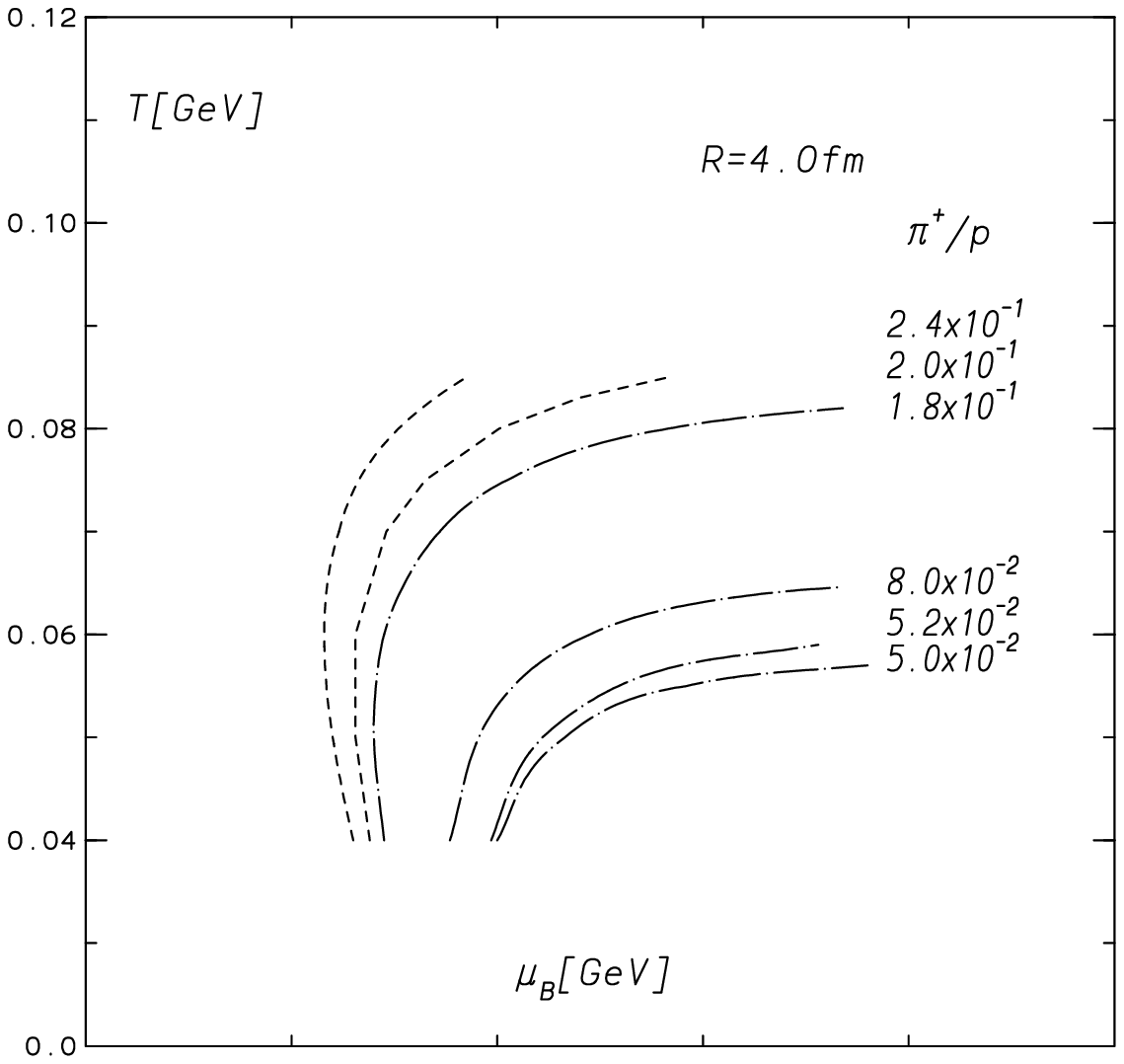,width=6.0 cm}\\
\vspace{-0.6 cm}
\epsfig{file=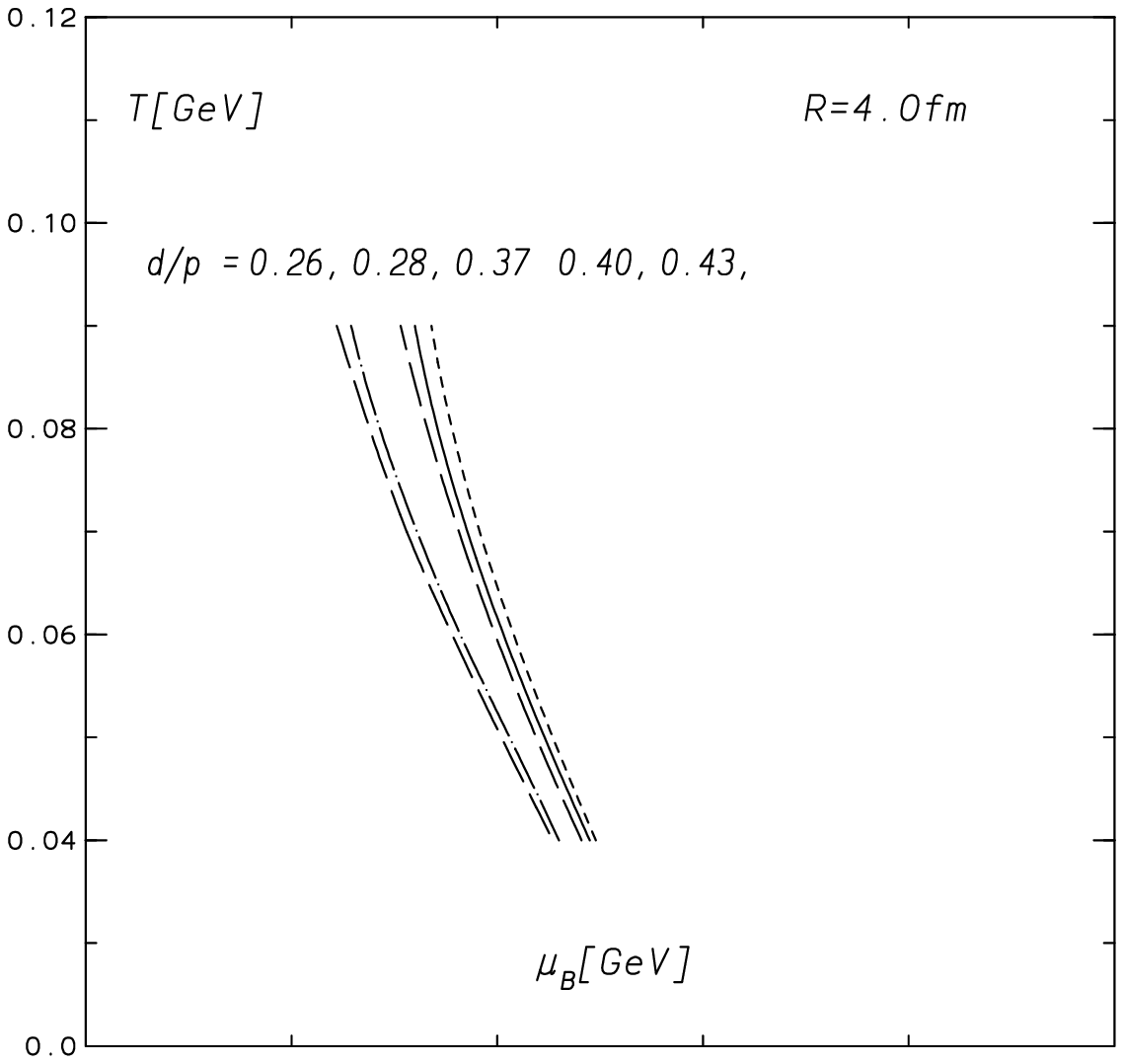,width=6.0 cm}\\
\vspace{-0.6 cm}
\epsfig{file=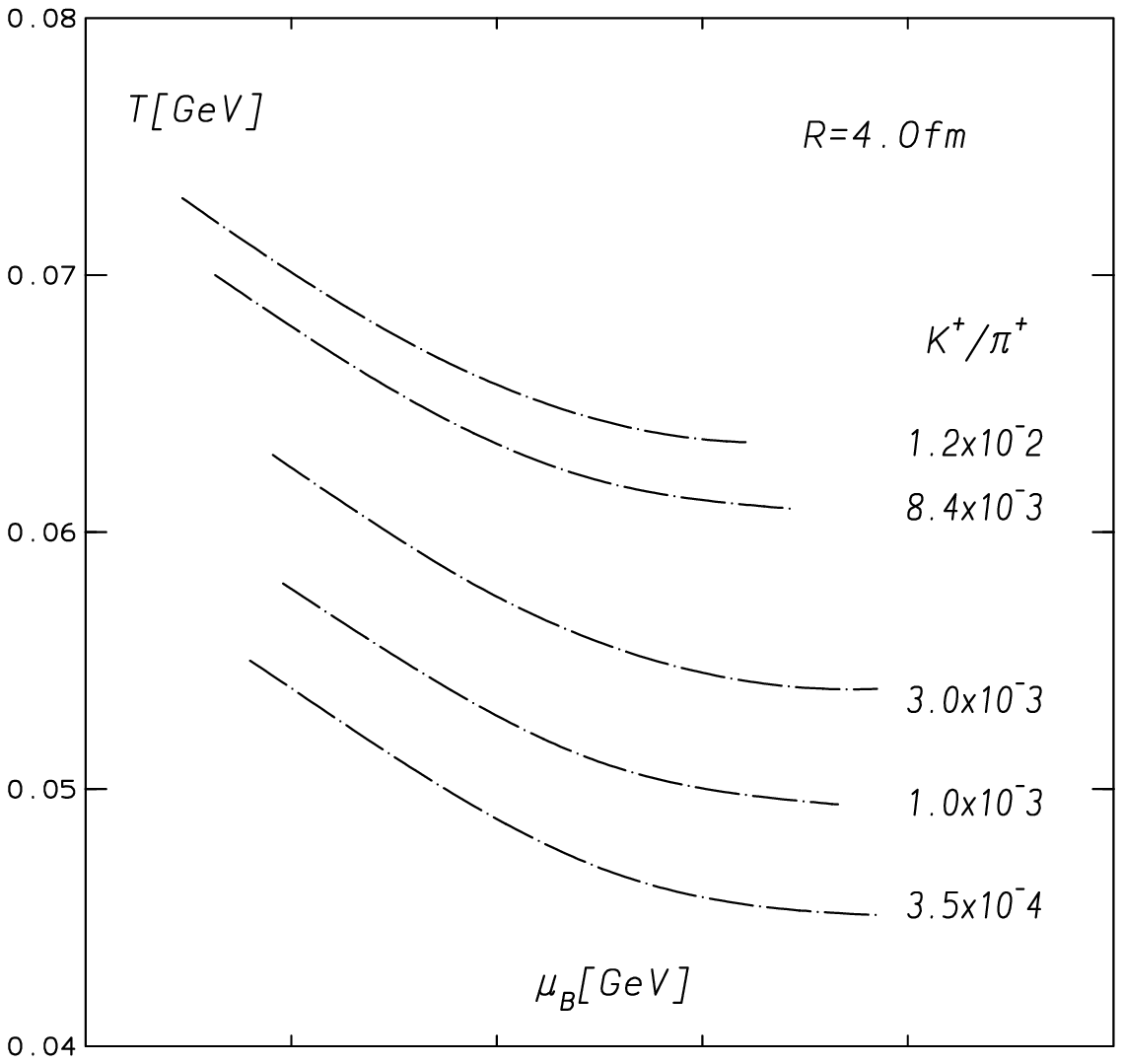,width=6.0 cm}\\
\vspace{-0.6cm}
\epsfig{file=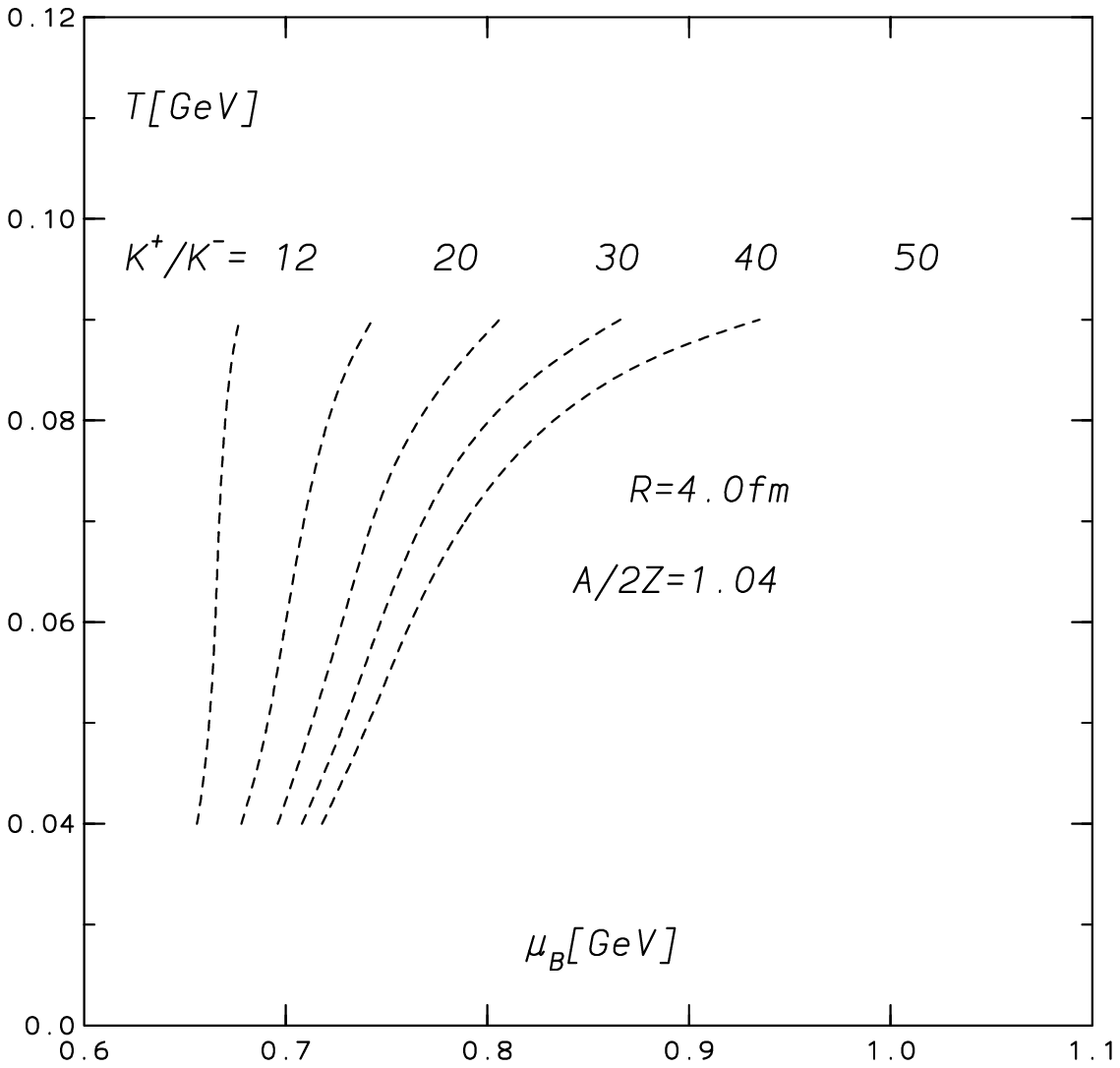,width=6.0 cm}\\
\end{center}
\end{figure}
\newpage

\begin{figure}
\begin{center}
\epsfig{file=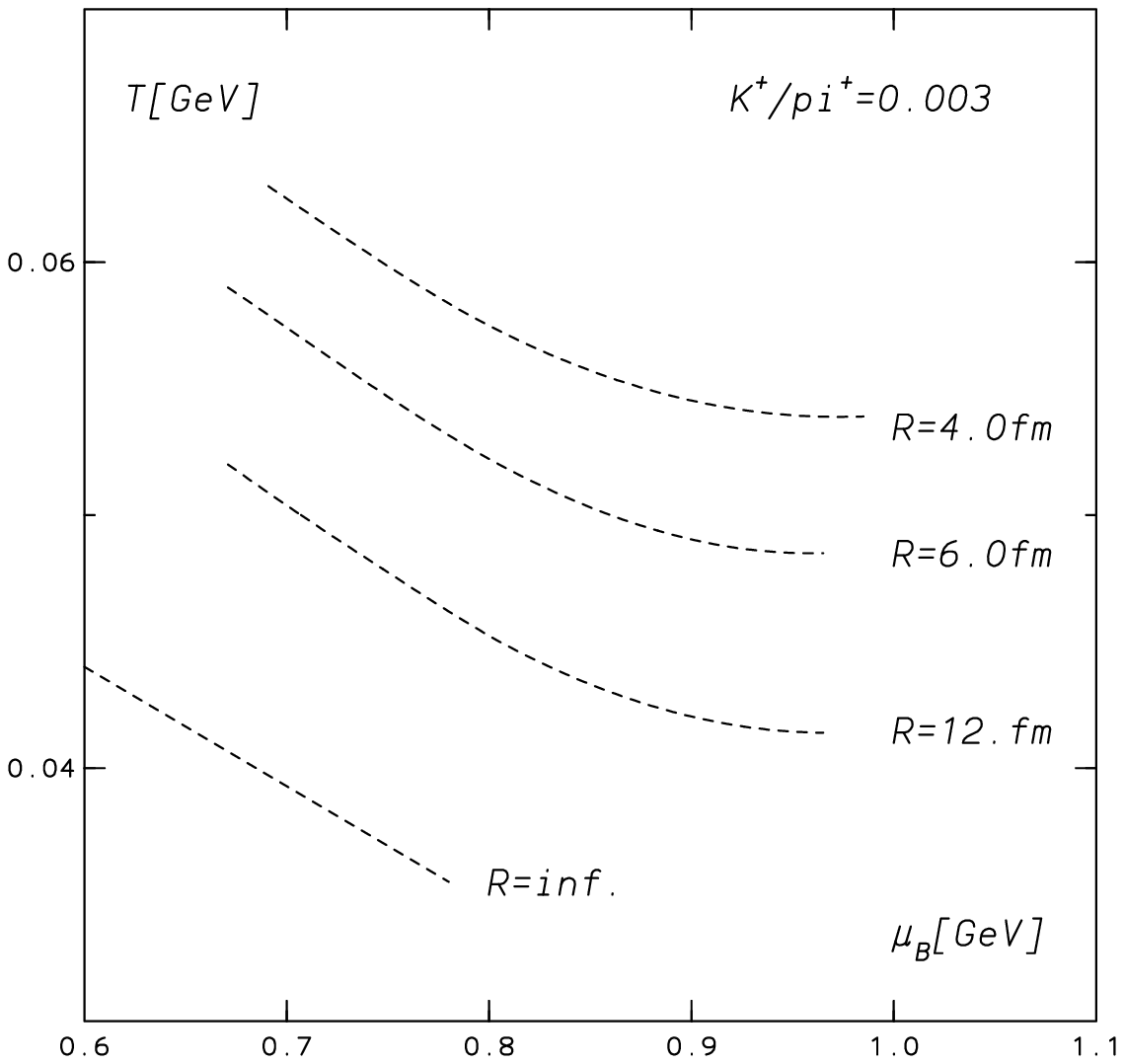,width=6.5cm}
\end{center}
\end{figure}
\newpage
\begin{figure}
\begin{center}
\epsfig{file=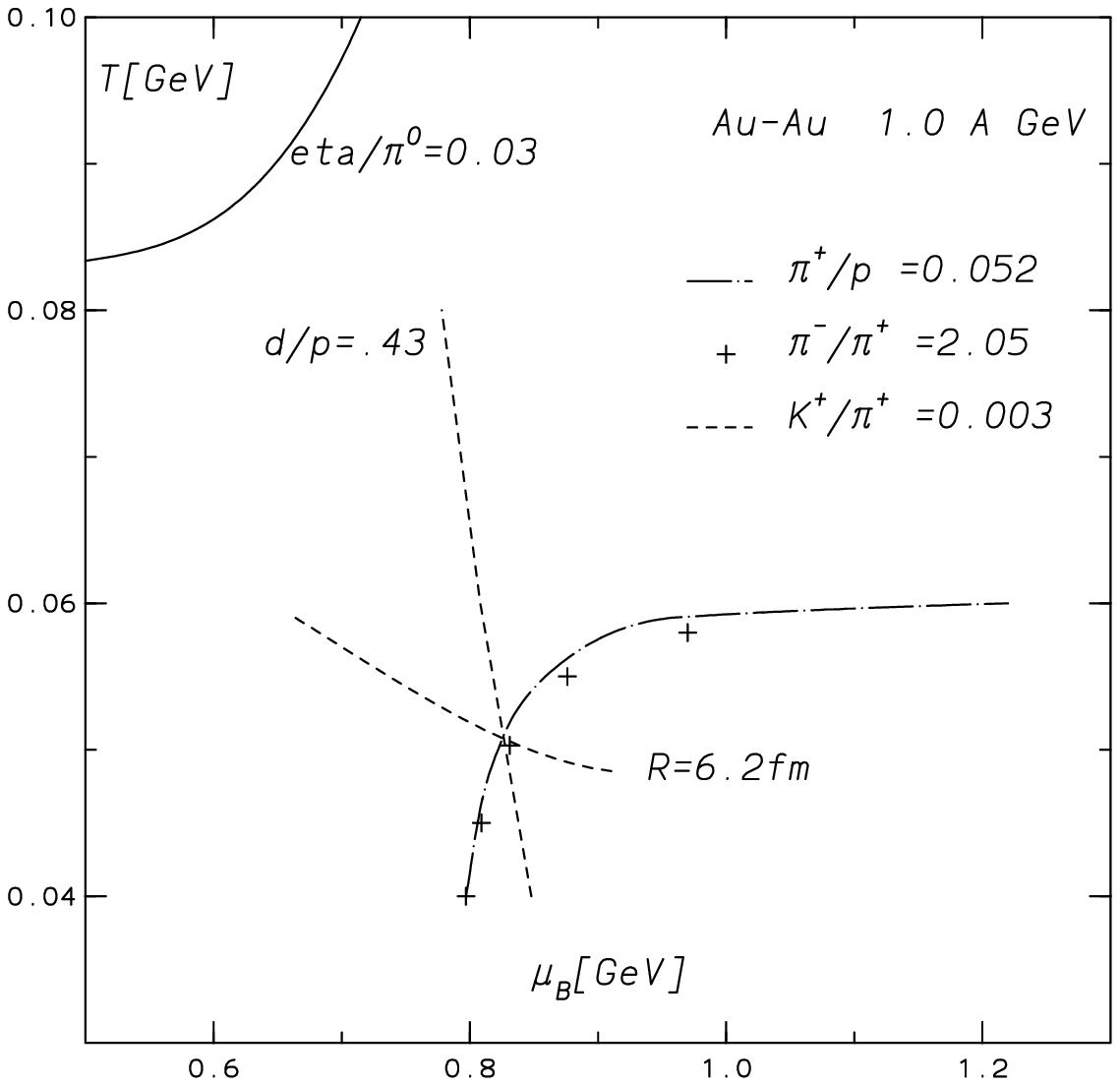,width=6.5cm}
\end{center}
\end{figure}
\newpage
%
%
%
%
\begin{figure}
\begin{center}
\epsfig{file=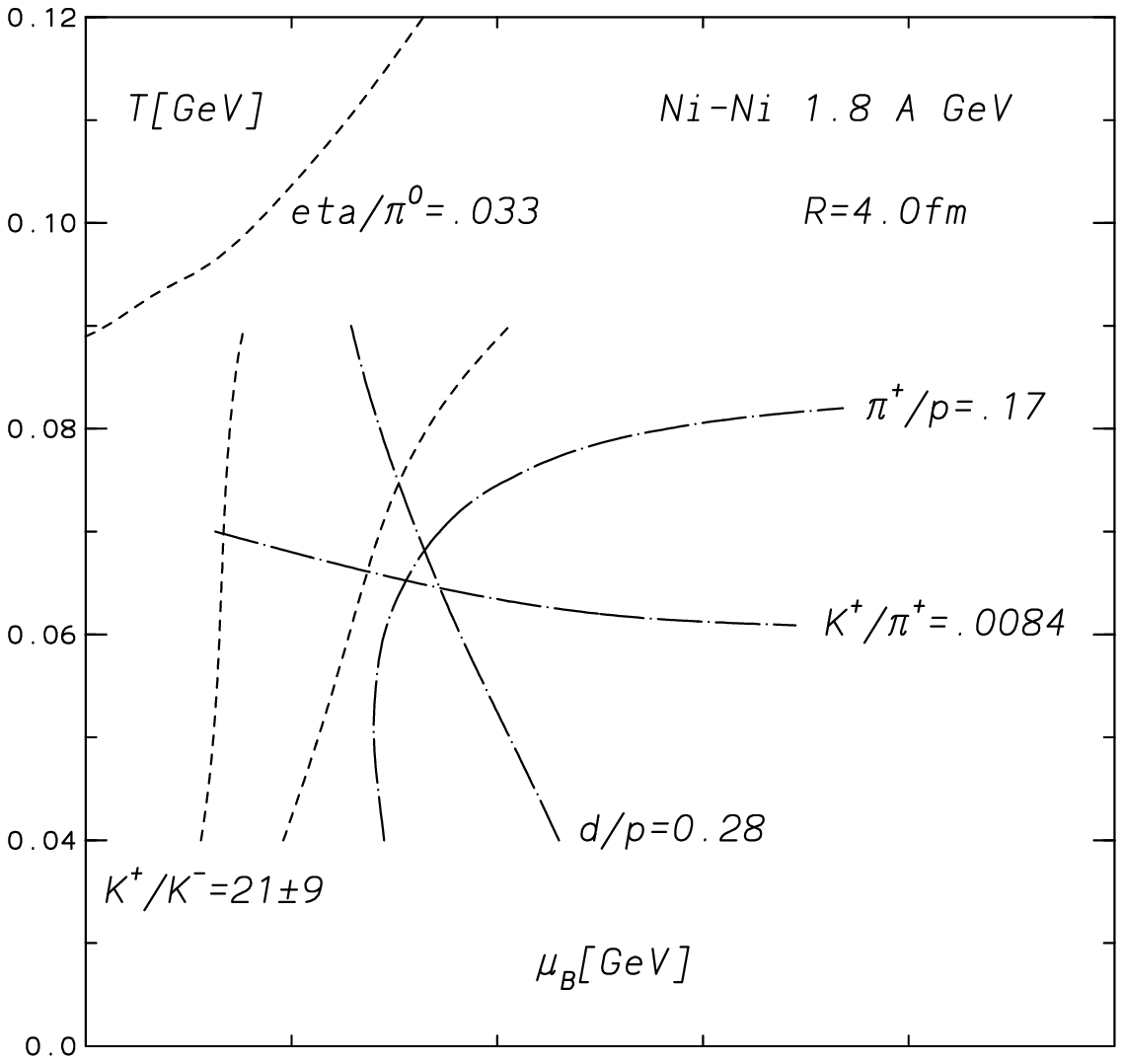,width=6.5 true cm}\\
\vspace{-0.6cm}
\epsfig{file=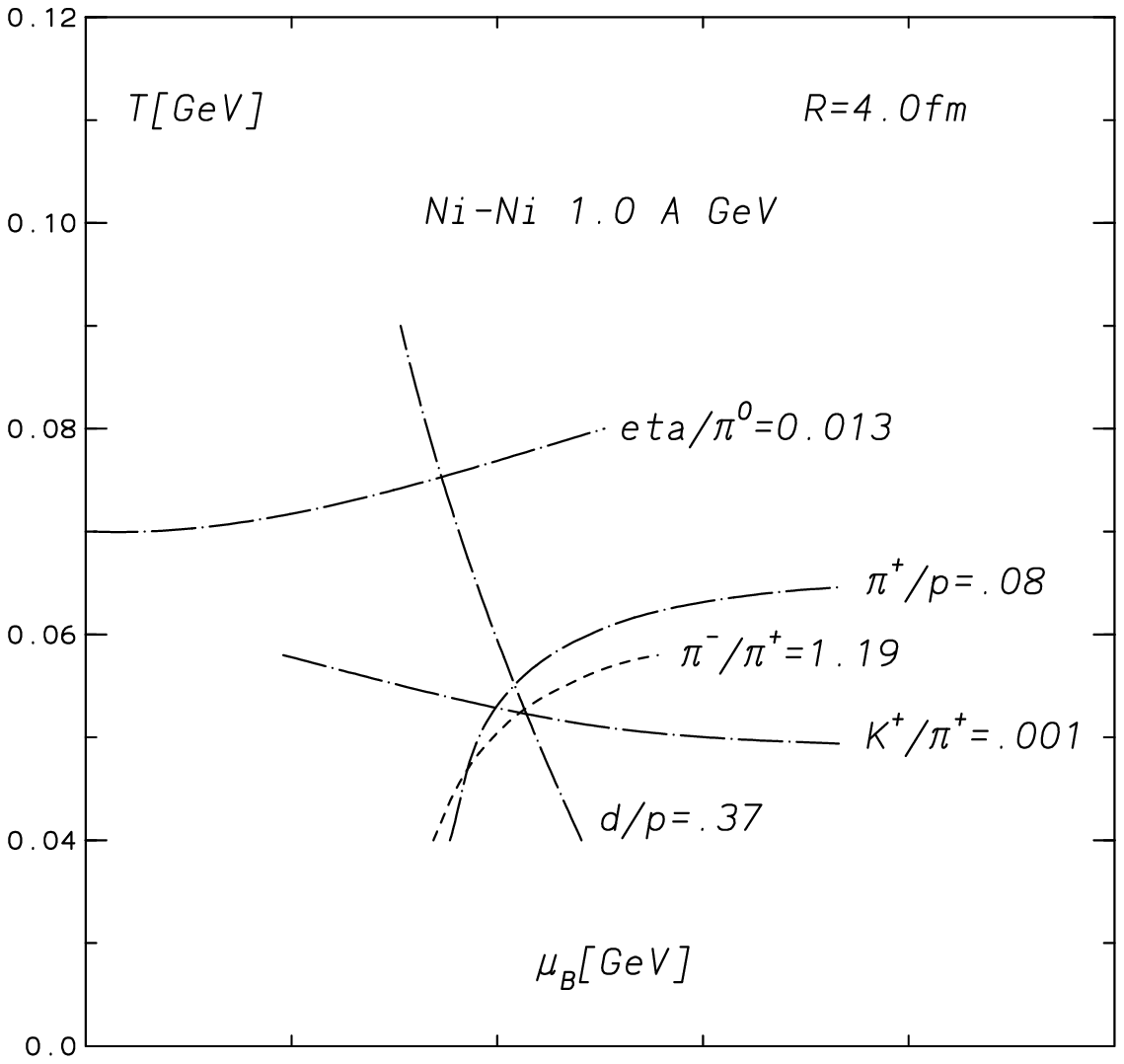,width=6.5 true cm}\\
\vspace{-0.6cm}
\epsfig{file=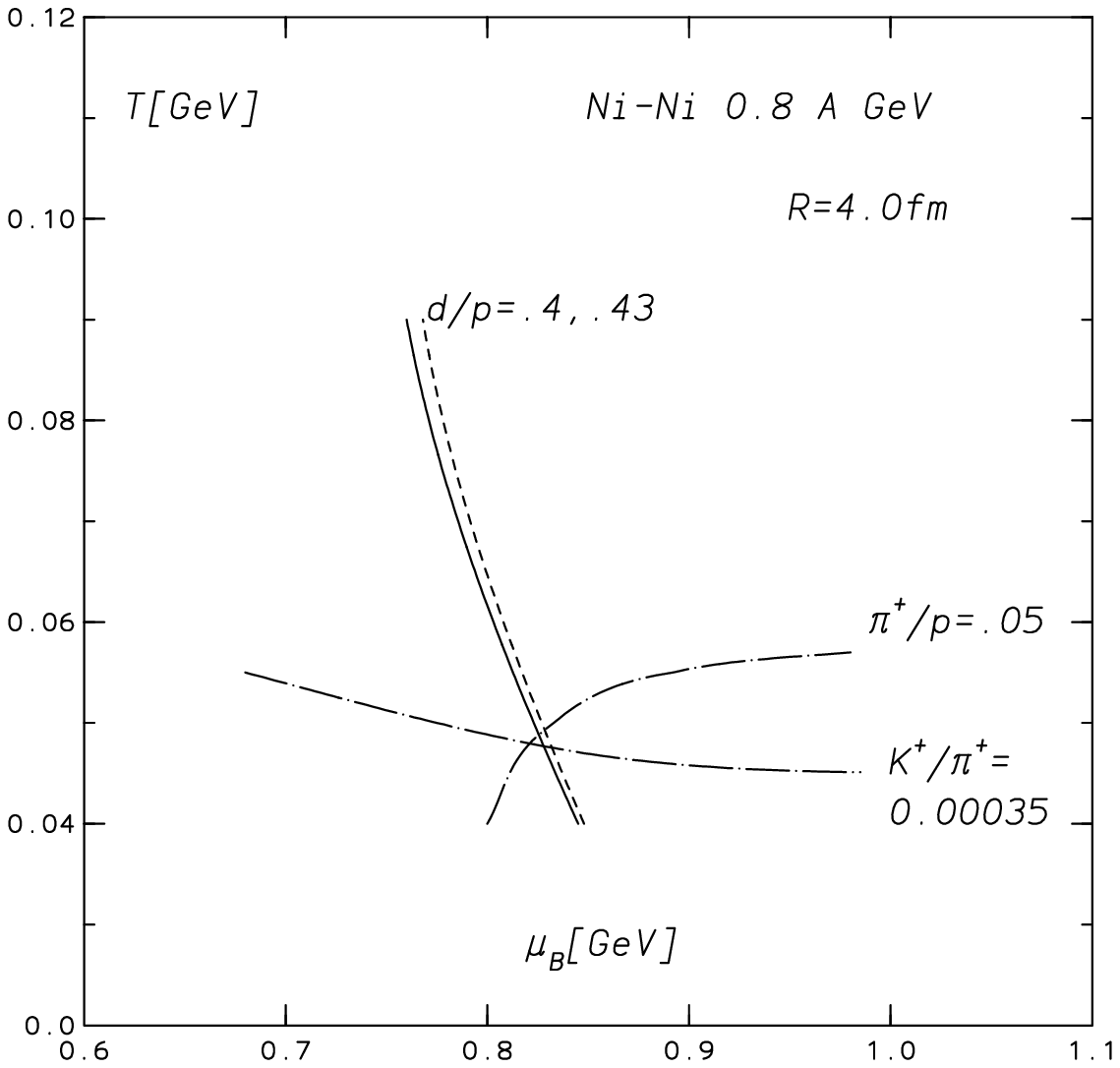,width=6.5 true cm}\\
\end{center}
\end{figure}
\newpage
%
%
%
%
%
%
\begin{figure}
\begin{center}
\epsfig{file=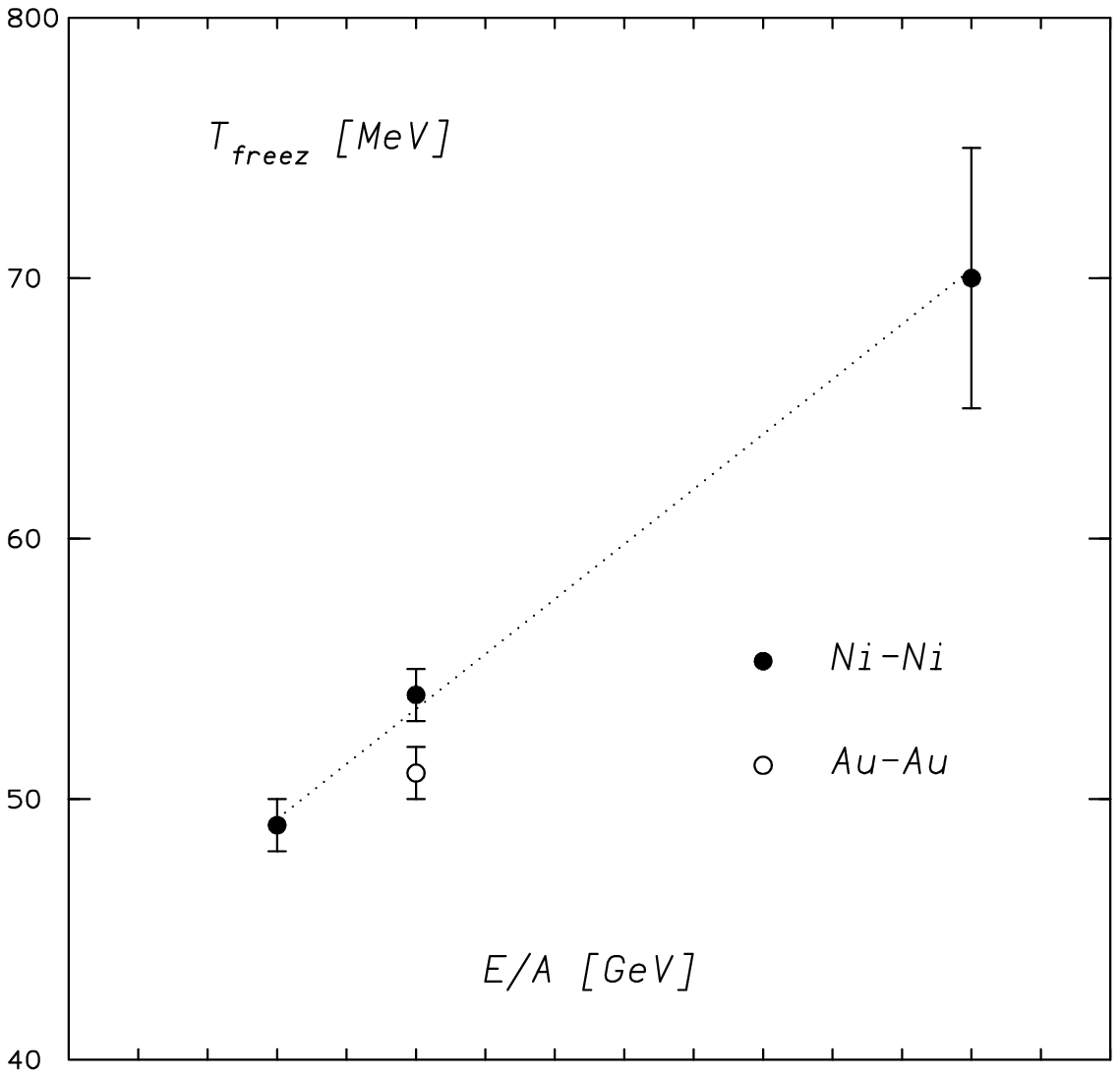,width=6.5cm}\\
\vspace{-0.6cm}
\epsfig{file=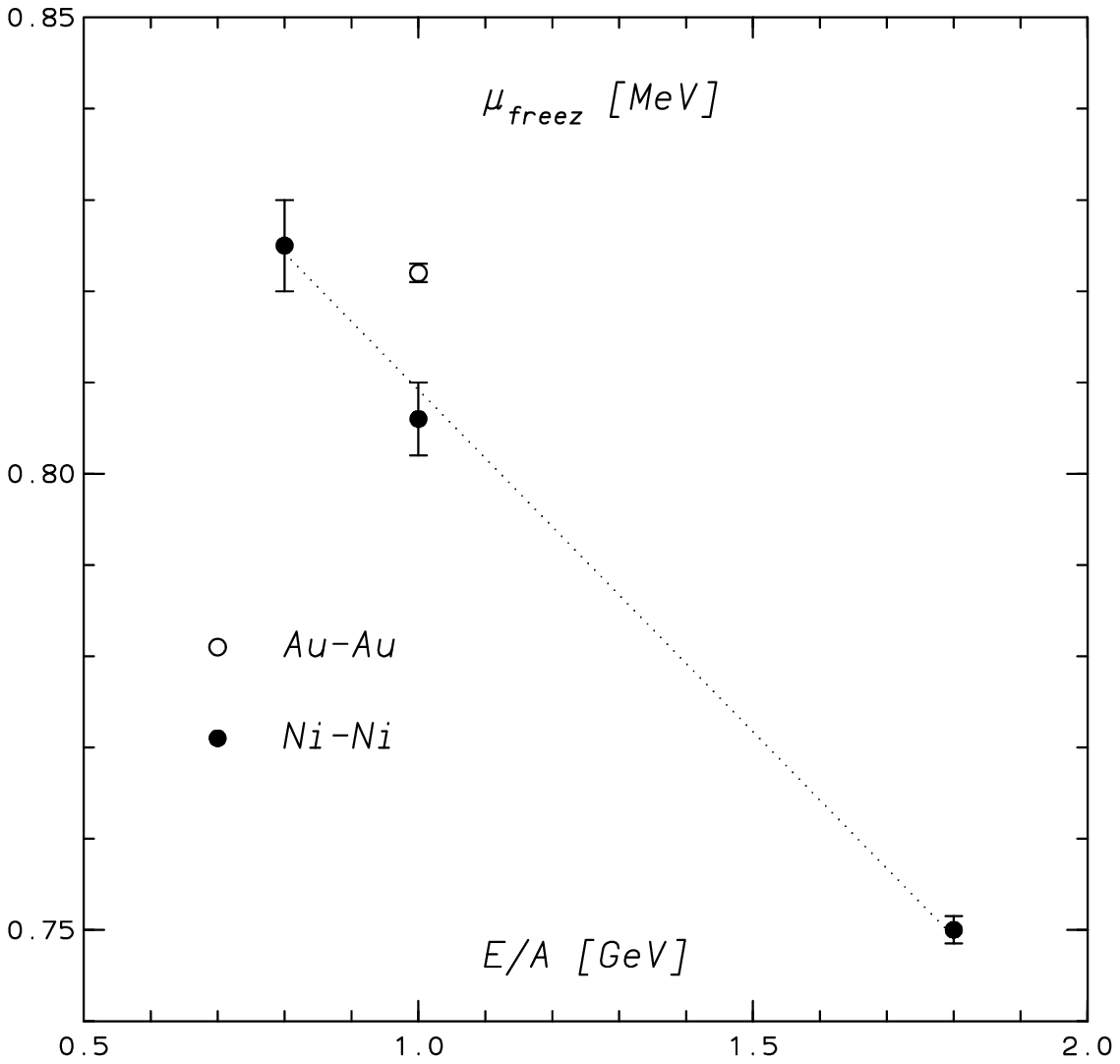,width=6.5cm}\\
\end{center}
\end{figure}
\newpage
\begin{figure}
\begin{center}
\epsfig{file=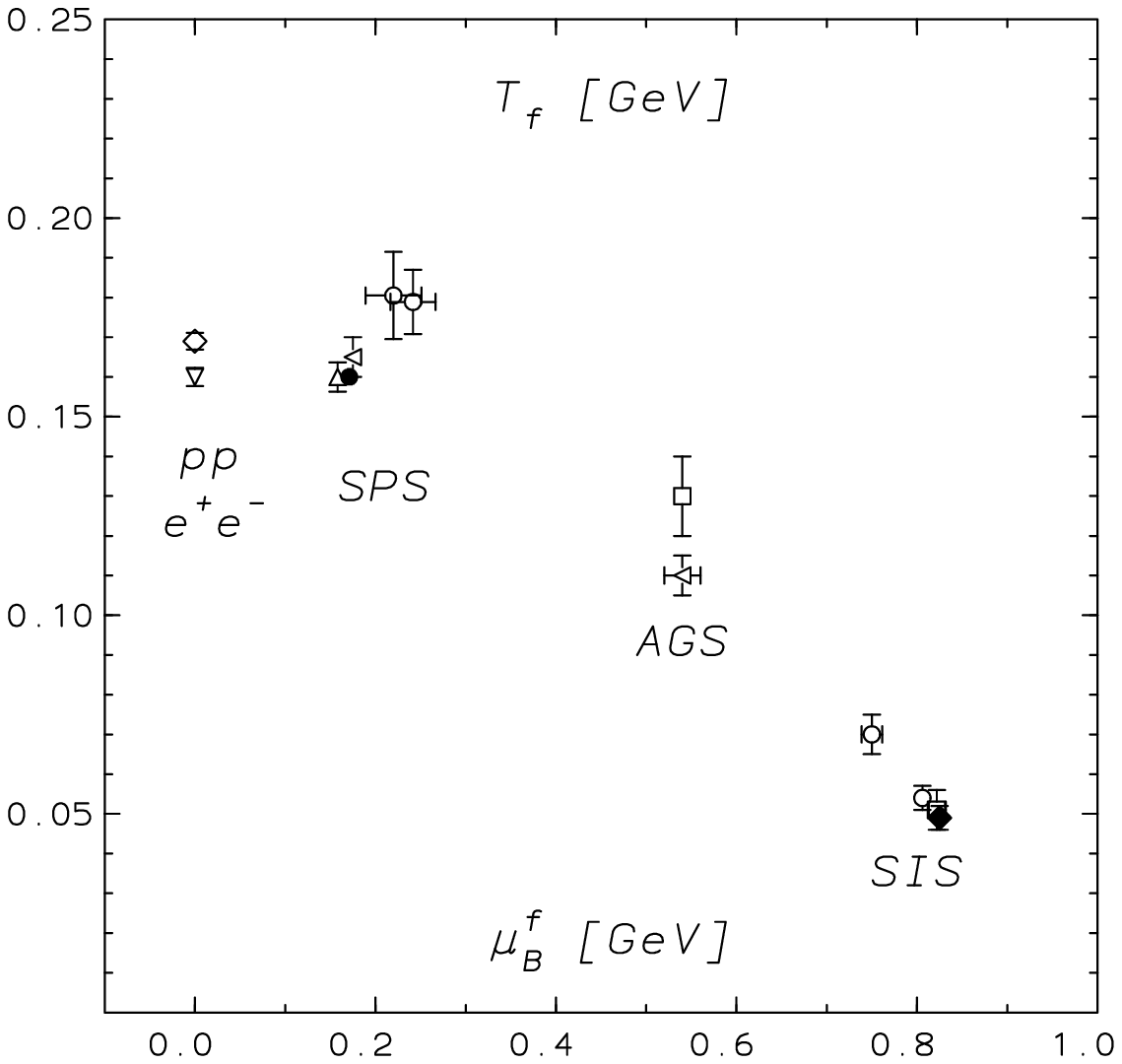,width=6.5cm}
\end{center}
\end{figure}
\newpage
\begin{figure}
\begin{center}
\epsfig{file=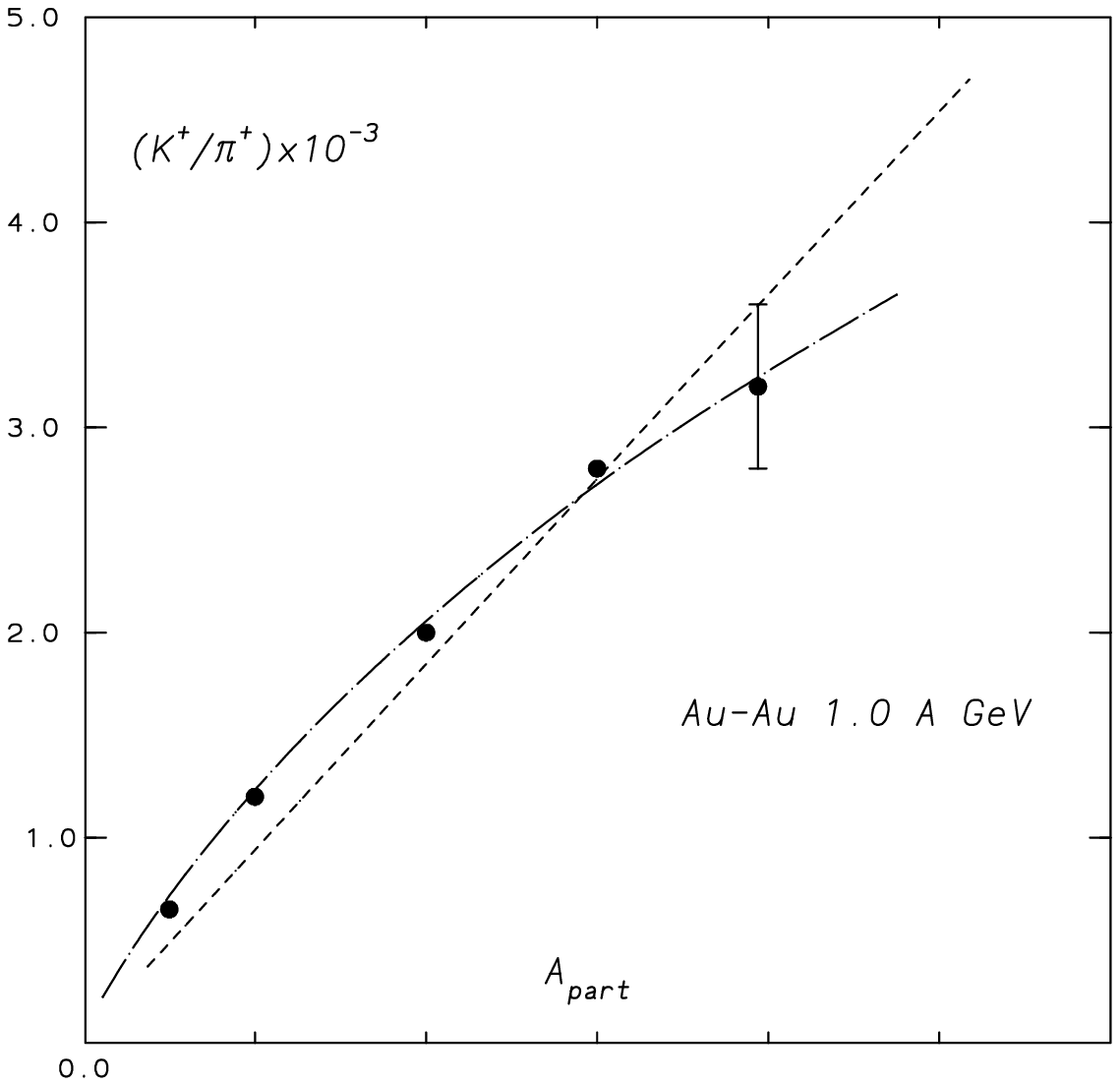,width=6.5cm}\\
 \vspace{-0.6cm}
 \epsfig{file=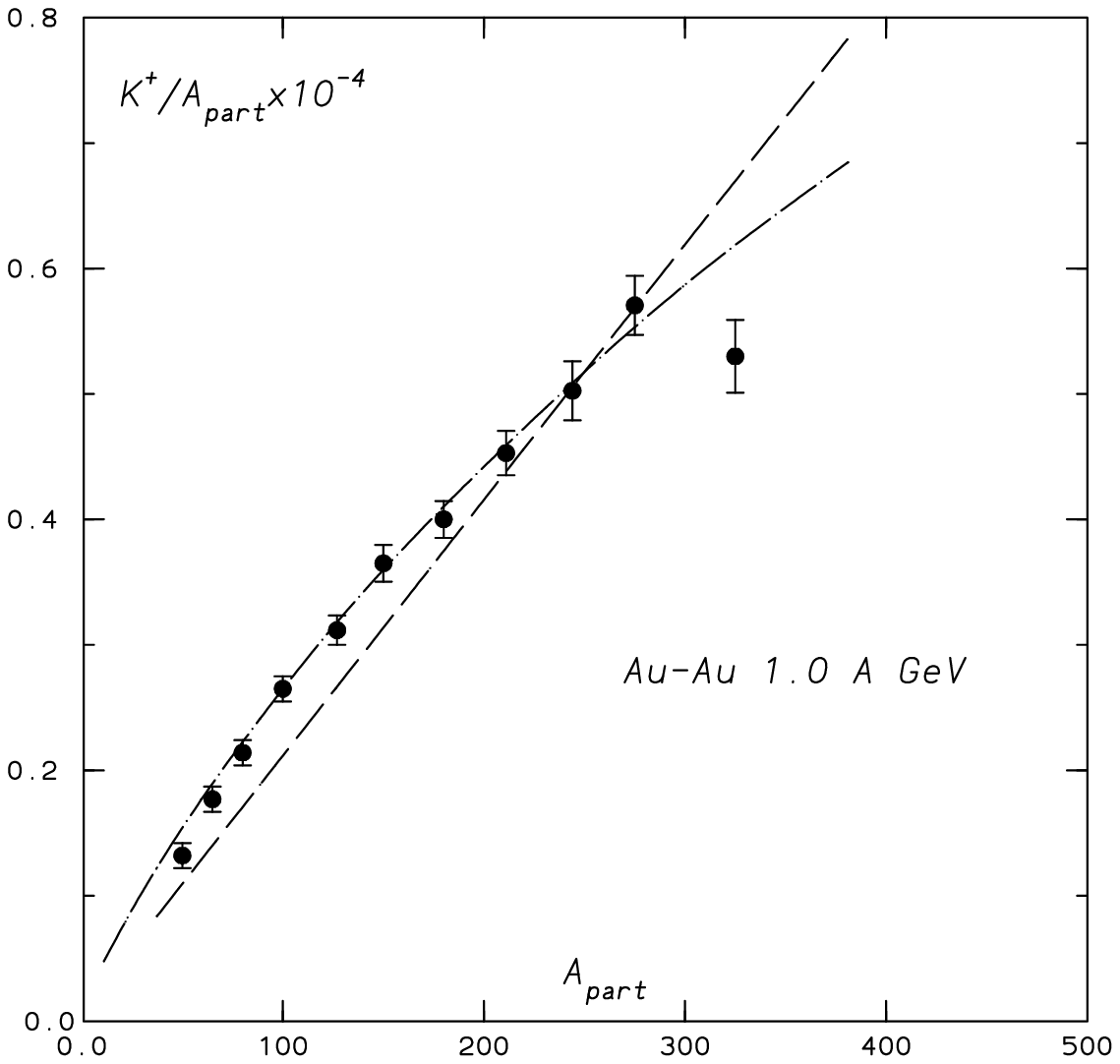,width=6.5cm}
\end{center}
\end{figure}
\newpage
%
%
%
%
\begin{figure}
\begin{center}
\epsfig{file=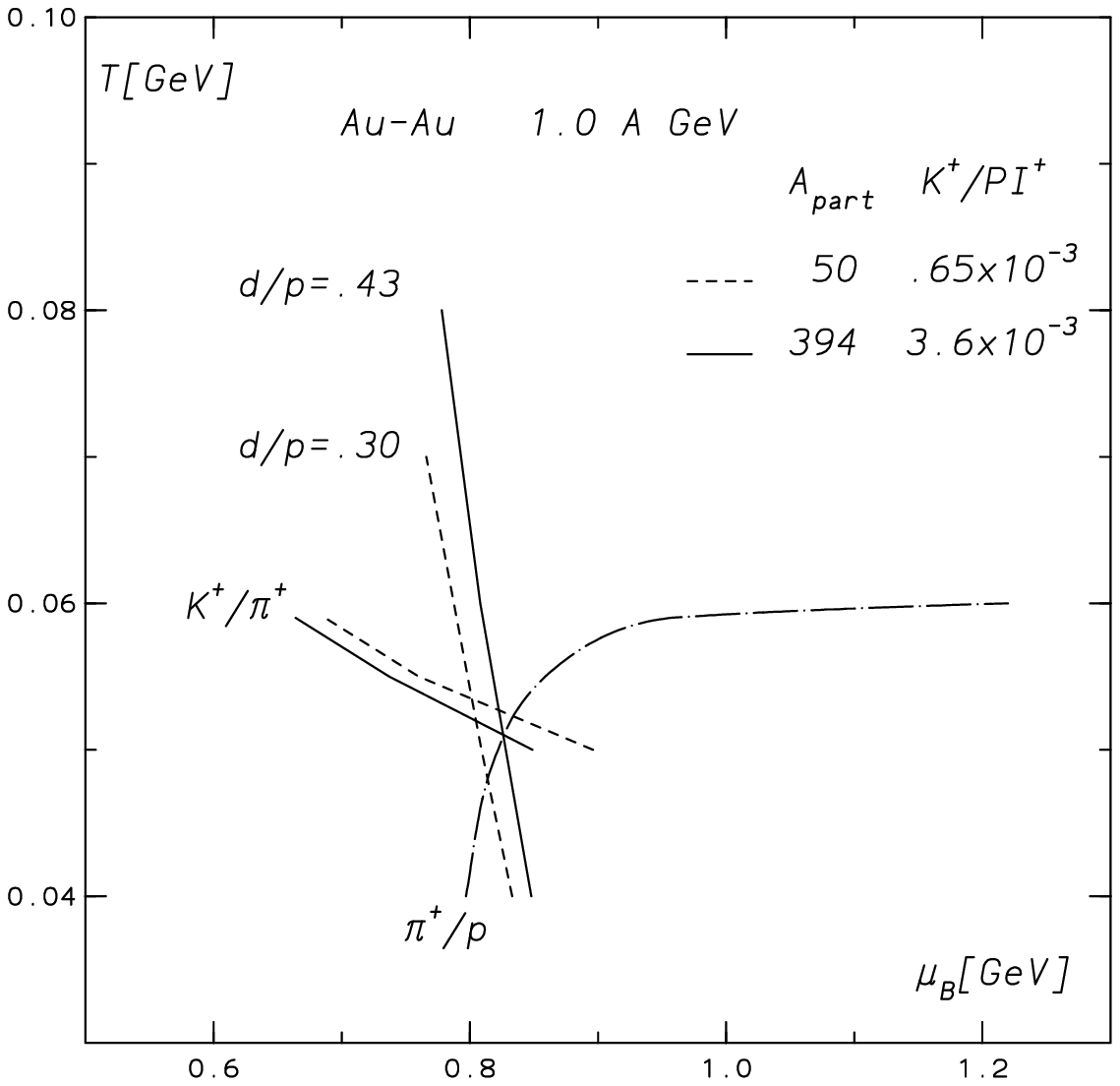,width=6.5cm}
\end{center}
\end{figure}
\newpage
%
%
%
%
\begin{figure}
\begin{center}
\epsfig{file=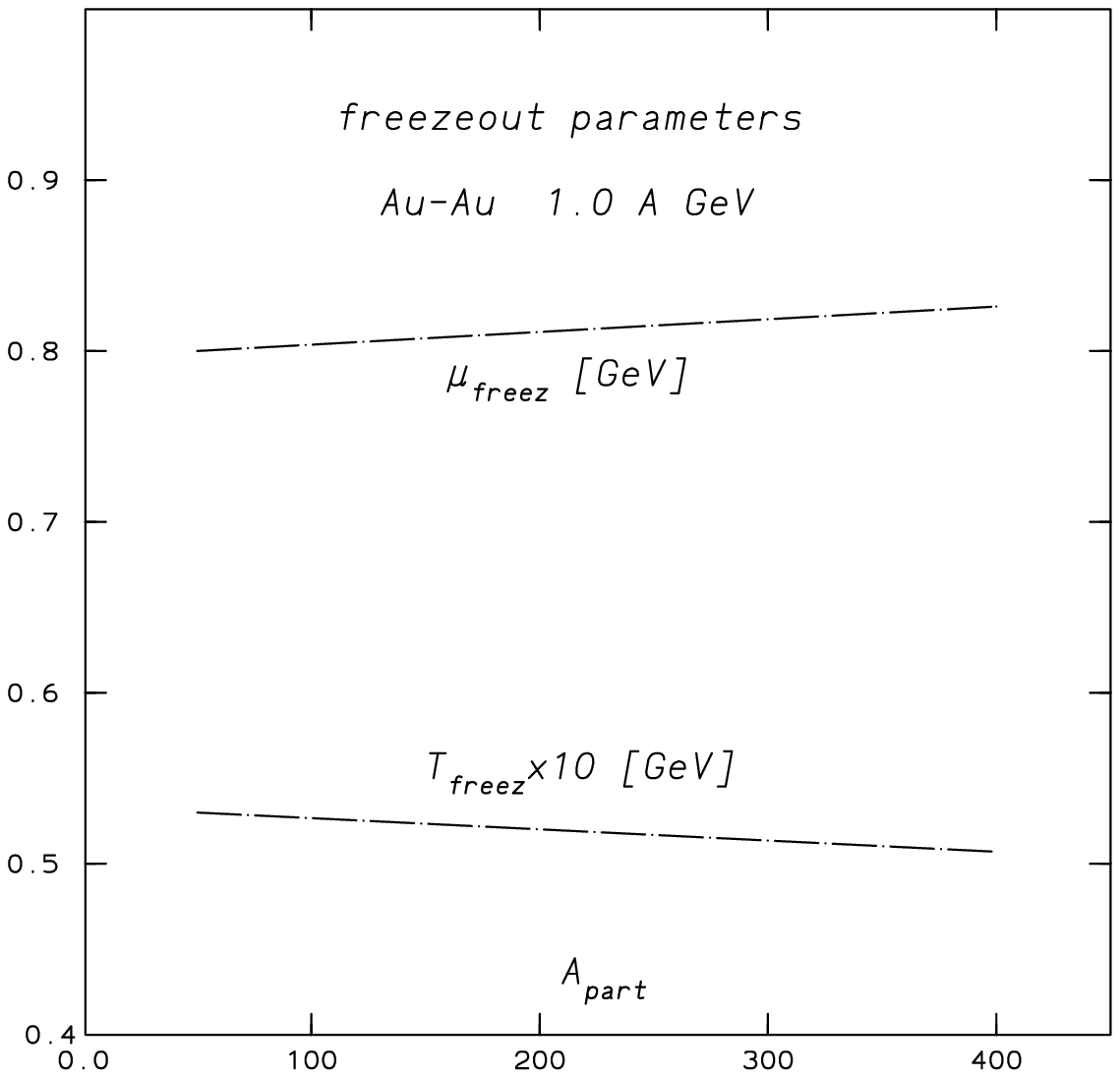,width=6.5cm}
\end{center}
\end{figure}
\newpage
\begin{figure}
\begin{center}
\epsfig{file=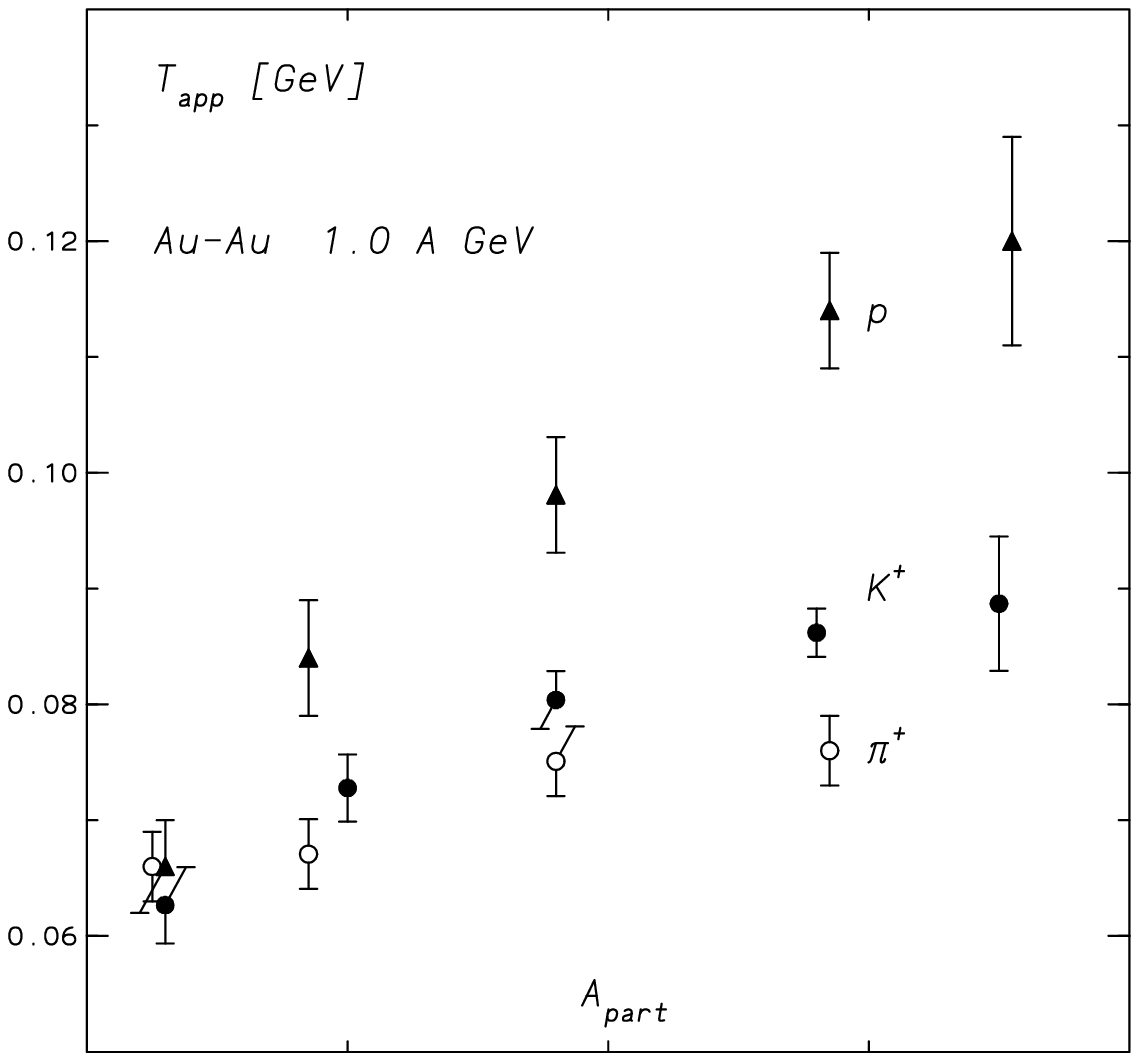,width=6.5 true cm}\\
\vspace{-0.6cm}
\epsfig{file=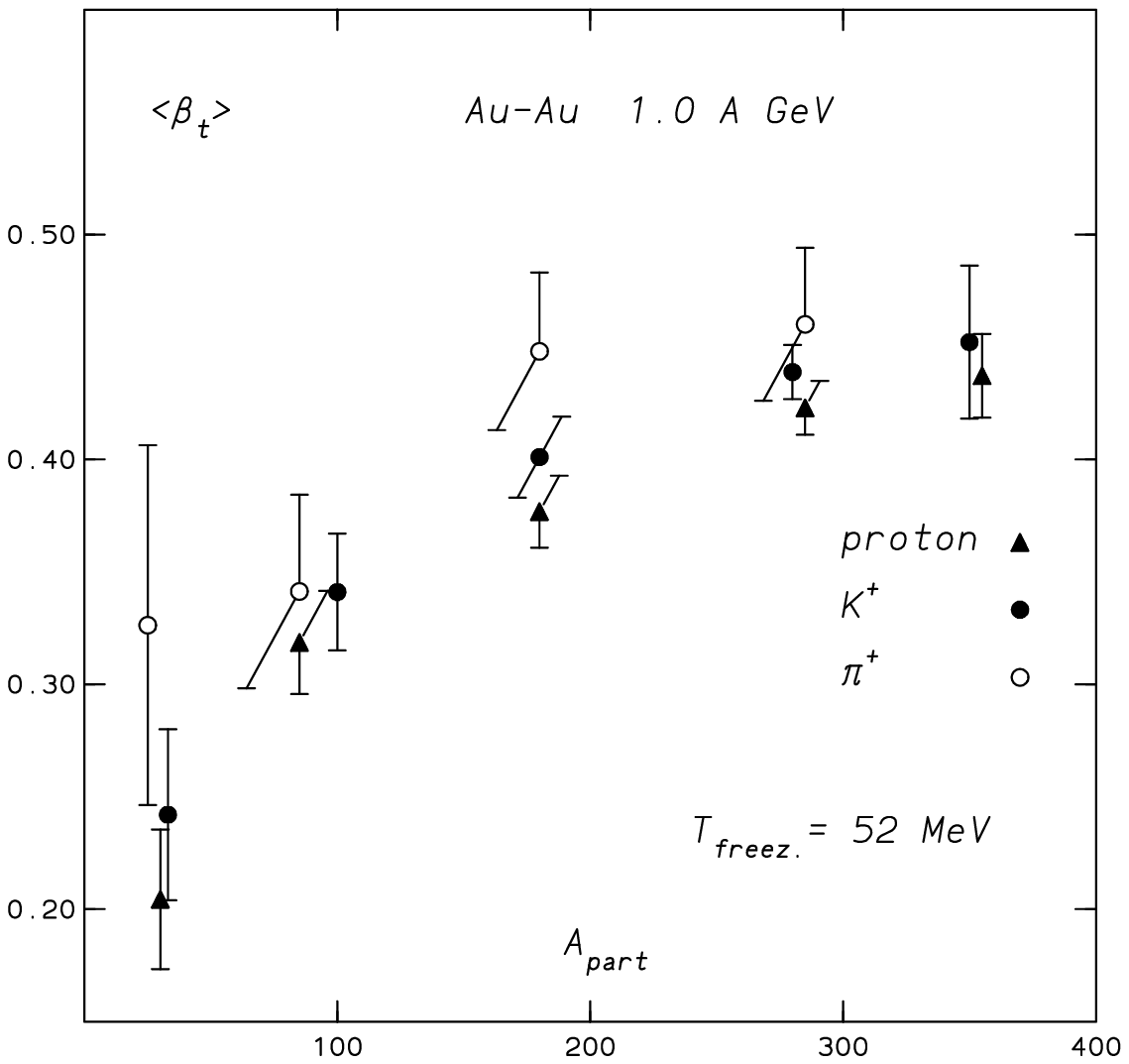,width=6.5 true cm}\\
\end{center}
\end{figure}

\section{Figure Caption}
Fig.1 {{\it Canonical strangeness suppression factor  (see text).}}\\

Fig.2. {{\it   Dependence  of  various  particle ratios
 on  the  radius
parameter $R$. }}\\

Fig.3 {{\it General trends for
  $\pi^+/p$, $d/p$,
 $K^+/\pi^+$ and
 $K^+/K^-$  freeze-out curve calculated
with $R=4.0$ fm and isospin asymmetry corresponding to $A/2Z=1.04$.}}\\

Fig.4 {{\it Dependence of $K^+/\pi^+$ freeze-out line with the
radius $R $ ..}}\\

Fig.5 {{\it Freeze-out lines corresponding to different particle ratios
measured in  Au+Au collisions at 1 A$\cdot$GeV.}}\\

Fig. 6 {{\it $T$ versus $\mu_B$ for central Ni+Ni collisions from 0.8 to 1.8
A$\cdot$GeV.}}\\

Fig.7    {{\it Freeze-out $T_f$ and $\mu_B^f$ as a function of incident energy
as extracted from the common crossing in
Figs.~5, 6 --
neglecting the results from $\eta/\pi^0$.}}\\

Fig.8    {{\it Freeze-out parameters $T_f$ and $\mu_B^f$ for SPS, AGS and SIS
energies.}}\\

Fig.9    {{\it Measured K$^+$/$\pi^+$ and K$^+$ multiplicity per $A_{part}$
 as a function of $A_{part}$
for Au+Au at 1 A$\cdot$GeV together with two calculations (see text).}}\\

Fig.10   {{\it Freeze-out parameters $T$ and $\mu_B$
for different $A_{part}$.}}\\

Fig.11   {{\it Variation of freeze-out parameters $T$ and $\mu_B$
as a function of $A_{part}$.}}\\

Fig.12   {{\it Apparent temperatures for $p,\pi^+ ,K^+$
(data from \cite{MANG97})
and calculated flow velocities
as a function of $A_{part}$
for Au+Au at 1 A$\cdot$GeV. }}\\

\end{document}